\documentclass[%
 reprint,
amsmath,amssymb,
aps,
pra,
]{revtex4-2}
\usepackage{graphicx}
\usepackage{dcolumn}
\newcommand{\cb}{\color{black}}
\newcommand{\cbl}{\color{black}}
\usepackage{bm}
\usepackage{xcolor}

\begin{document}

\preprint{APS/123-QED}

\title{Quantum Excitation Transfer in an Artificial Photosynthetic Light-Harvesting System}

\author{Stephon Alexander$^1$}
\author{Roger Andrews$^2$}
\author{Oliver Fox$^3$}
\author{Sarben Sarkar$^4$}
\affiliation{$^1$Brown Theoretical Physics Center and Department of Physics, Brown University, Providence, Rhode Island 02912, USA}
\affiliation{$^2$Department of Physics, The University of the West Indies, St. Augustine, Trinidad and Tobago}%
\affiliation{$^3$Department of Physics and Astronomy, University of Exeter, Exeter EX4 4QL, United Kingdom}
\affiliation{$^4$Department of Physics, King’s College London, Strand, London WC2R 2LS, United Kingdom}

\date{\today}

\begin{abstract}
We analytically derive transfer probabilities and efficiencies for an artificial light-harvesting photosynthetic system, which consists of a ring coupled to a central acceptor. For an incident photon pair, we find near-perfect single excitation transfer efficiency with negligible double excitation transfer in the weak coupling regime. In the strong coupling regime, single excitation transfer efficiency was greater than $90\%$, while the double excitation efficiency was approximately $50\%$. We have found that the three main factors which determine high transfer efficiencies are large acceptor probabilities, long acceptor decay times, and strong photon-ring coupling. A possible implementation of the theoretical framework to bio-inspired solar energy devices is also discussed.
\end{abstract}

\maketitle

\section{\label{sec:level1}Introduction}
Light-harvesting (LH) complexes consist of pigment-proteins that capture sunlight and transport it to a reaction center (RC) for photosynthesis~\cite{bio1,bio2,bio3,bio4,Harel2012}. Specifically, LH1 systems found in purple photosynthetic bacteria generally consist of protein complexes which form a ring around a RC~\cite{bio5,biolh1,biolh2}. These LH1-RC systems are tightly packed and organized in a way to give high photon transfer efficiencies and can differ across species under different light conditions~\cite{lhmodel1}.

The quantum nature of the LH process has been an important area of research~\cite{Herek2002,Ritz2002,Curutchet2017} and such systems are often modeled through quantum Hamiltonians~\cite{Ritz1998,wykeroger,lhtransportnoise,Tan2012,Chuang2020,Andrews2022}. Ritz \textit{et al.}~\cite{Ritz1998} established an effective Hamiltonian of the circular bacterial photosynthetic unit and determined one- and two-exciton spectra. Wyke \textit{et al.}~\cite{wykeroger} modeled the antenna system as a ring of 2-level systems (2-LSs) and found near-perfect transfer efficiency for a single photon and a reduced efficiency with a laser pulse. Caruso~\textit{et al.}~\cite{lhtransportnoise} examined the effect of noise in excitation transport of LH systems and identified mechanisms for additional transport channels due to dephasing. Tan and Kuang~\cite{Tan2012} investigated quantum phase transitions in environment-assisted LH systems via a Lindblad master equation, and found high efficiency close to the critical point. Chuang and Brumer~\cite{Chuang2020} examined steady state incoherent light-matter interaction conditions in an LH1-RC complex to more effectively model the natural environment. Andrews \textit{et al.}~\cite{Andrews2022} used a Lindblad master equation and demonstrated numerically that disorder in the intra-ring couplings had a negligible effect on transfer efficiency, however photon-ring disorder caused a significant decrease close to resonance.

Other quantum effects in LH systems have also been investigated~\cite{lhentang,lhcoherence,Dongdarkstate}. Sarovar \textit{et al.}~\cite{lhentang} examined a framework to describe entanglement in LH complexes, and applied it to a specific system to show the existence of long-range and multipartite entanglement. Str\"umpfer~\cite{lhcoherence} investigated the role of quantum coherence and determined the effect of energy level shifts and resonances for increasing energy transfer rates. Dong \textit{et al.}~\cite{Dongdarkstate} examined a ring of coupled 2-LSs and demonstrated that the collective ring can be modeled as a $\Lambda$-type 3-level system (3-LS). In addition, they determined the efficiency of a ring coupled to a photon and a central two-level RC. They explained the high transfer efficiencies and power outputs using dark state channels.

Two-photon transfer in many systems has been an area of active research~\cite{Aiyejina2024,Dong2012,alexanian,corrtwop,Hardal2014,russo}. Aiyejina \textit{et al.}~\cite{Aiyejina2024} considered a trimer of 3-LSs and found conditions for perfect and near-perfect double excitation transfer for laser pulses and single photons. Dong \textit{et al.}~\cite{Dong2012} considered coupled cavity arrays containing 3-LSs and determined that uniform intercavity coupling was most applicable for quantum-state transfer. Alexanian~\cite{alexanian} considered two-photon exchange in a cavity containing two 3-LSs, Rabi oscillations and entanglement for symmetric and antisymmetric states. Liao and Law~\cite{corrtwop} explored two photon transport properties in a cavity with a nonlinear medium and demonstrated the emergence of quantum correlations between the two photons. In a two nonlinear cavity system, Hardal and M\"ustecaplioglu~\cite{Hardal2014} considered coherences and entanglement to find an advantage of practical realization of entanglement for two-photon over one-photon exchange. Russo \textit{et al.}~\cite{russo} identified photon pair hopping between two cavities separated by a vibrating mirror.

Artificial light-harvesting systems have been experimentally realized, such as the LH1-RC photosynthetic complex~\cite{exp1,exp3,exp4}. Suemori et al.~\cite{exp1} isolated LH1-RC units onto ITO electrodes to demonstrate the methodology that can be used to build artificial photosynthetic units. Photocurrent response was measured, consistent with the function of capturing light and transferring electrons. Sumino et al.~\cite{exp3} designed assemblies of LH1-RC arrays placed on domain-structured planar lipid bilayers. Intermolecular energy transfer to the LH1-RC via LH2 was observed. Rousseaux et al.~\cite{exp4} fabricated a Russian doll complex with two poryphin nanorings which reflect LH1-RC architecture. Photophysical experiments demonstrated excitation transfer between the two rings. Recent advancements have also demonstrated the use of photosynthetic and artificial photosynthetic systems in order to harvest solar energy~\cite{Nkinyam2025,Machn2023,Gust2009}.

In pursuit of breakthroughs in renewable energy and sustainable technologies, we present an analytical investigation into the probabilities and transfer efficiencies within an artificial light-harvesting photosynthetic system engineered to replicate key features of natural energy conversion. This study models a donor ring coupled to a central acceptor as a basis for developing next-generation solar energy devices. Our analysis reveals that, for an incident photon pair, the system achieves near-perfect single excitation transfer in the weak coupling limit, while in the strong coupling regime, single excitation transfer efficiencies exceed $90\%$ and double excitation efficiencies reach approximately $50\%$. These findings underscore three pivotal design criteria that can be strategically exploited to optimize device performance: maximizing the acceptor’s effective probability, prolonging its decay time, and enhancing the photon-ring coupling. Ultimately, this work not only advances our theoretical understanding of excitation dynamics but also provides a theoretical framework for the practical development of highly efficient, bio-inspired light-harvesting technologies.

In this paper we analytically determine transfer efficiencies of two LH1-RC models which consist of a ring of 3-LSs coupled separately to a 2-level and a 3-level RC. In Sec.~II we introduce the LH quantum system framework and derive analytic transfer probabilities and efficiencies for a range of collective, localized and delocalized states. We expand the model to include coupling to photon pairs, finding analytic solutions with perturbation theory. In Sec.~III we present results and analyses of probabilities and efficiencies for a range of system parameters. Finally, a conclusion is given in Sec.~IV.

\section{\label{sec:level2}Theory}
\begin{figure}[t]
\includegraphics[width=0.48\textwidth]{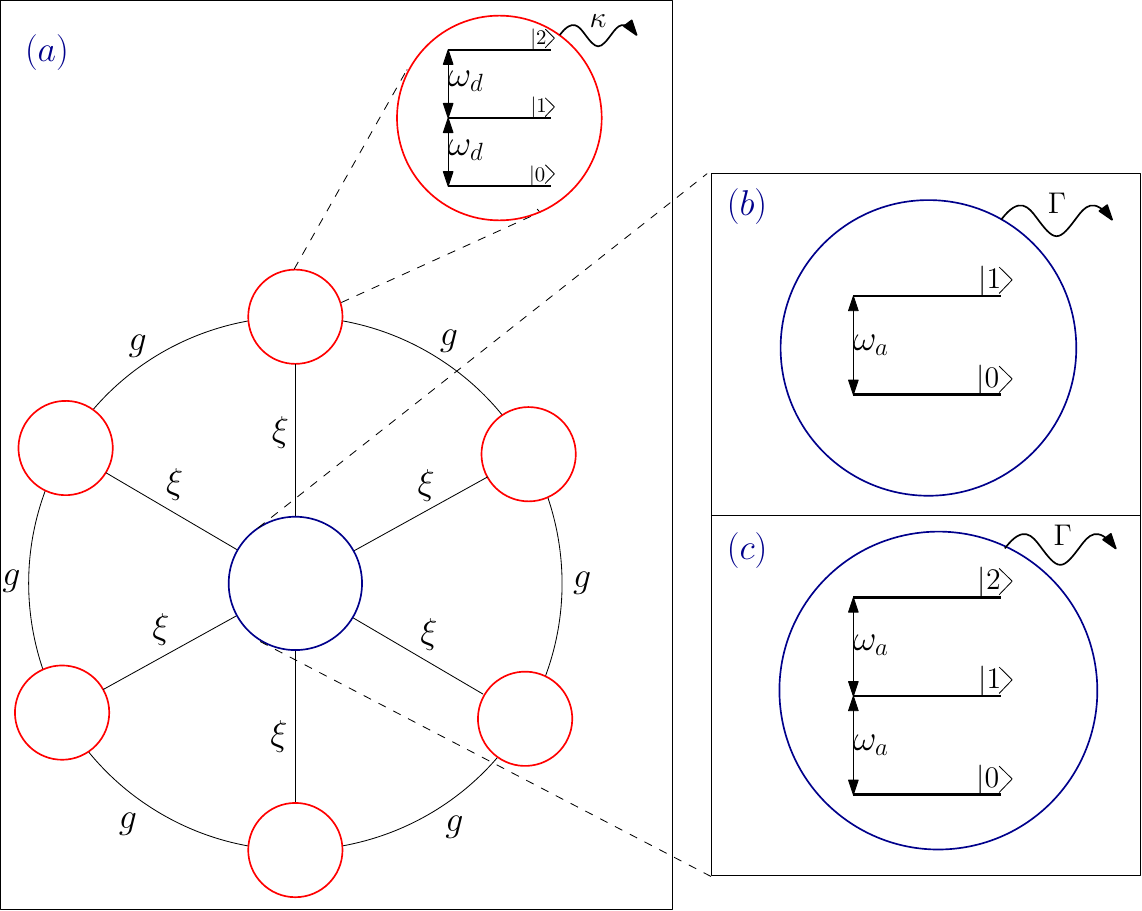}
\caption{\label{fig:pic}(a) Schematic showing the donor ring, made up of 3-LSs, each in the ladder configuration, with allowed transition energies $\omega_d$. The coupling constant between the nearest-neighbor donor 3-LSs is $g$ and between each donor 3-LS and the acceptor is $\xi$. The acceptor is modeled as (b) a 2-LS with transition energy $\omega_a$ and (c) a 3-LS with two allowed transitions each with energy $\omega_a$, and decay rate $\Gamma$.}
\end{figure}
The LH1 antenna system is made up of $N$ 3-LSs, each in the ladder configuration, coupled with their nearest-neighbors to create a ring structure (see Fig.~\ref{fig:pic}(a)). The Hamiltonian of the ring system, $H_d$ is given as
\begin{equation}
\label{eq:ringhamil}
    H_d=\sum_{j=1}^N\left[(\omega_d-i\kappa)e_j^\dagger e_j+g(e^\dagger_je_{j+1}+h.c.)\right],
\end{equation}
where the creation operator $e_j^\dagger$ is defined as (see Appendix~\ref{appen1})
\begin{align}
\label{eq:ejdef}
    e^\dagger_j=|1_j\rangle\langle0_j|+\sqrt{2}|2_j\rangle\langle1_j|,
\end{align}
for $j\in\{1,N\}$, that act on the $j^\text{th}$ 3-LS. The $j^\text{th}$ 3-LS has states $|0_j\rangle, |1_j\rangle$ and $|2_j\rangle$ with corresponding energies $0, \omega_d$ and $2\omega_d$. The coupling constant between the nearest-neighbor donor 3-LSs is $g$. Spontaneous emission is described phenomenologically using the decay constant $\kappa$.

\subsection{\label{sec:sea}LH1 Ring Coupled to a 2-LS Acceptor without Light}
The 2-LS acceptor is coupled to a RC with decay rate $\Gamma$ (see Fig.~\ref{fig:pic}(b)). It has energy level spacing $\omega_a$ with Hamiltonian $H_a$ given as
\begin{equation}
    H_a=(\omega_a-i\Gamma)a^\dagger a,
\end{equation}
where the creation (annihilation) operator $a^\dagger$ ($a$) is defined as $|1_a\rangle\langle0_a|$ ($|0_a\rangle\langle1_a|$). 

The donor-acceptor interaction Hamiltonian $H_{da}$, is defined as
\begin{equation}
\label{eq:hda}
    H_{da}=\xi\sum_{j=1}^N(e_j^\dagger a+e_j a^\dagger),
\end{equation}
where each donor 3-LS is coupled to the 2-LS acceptor with coupling constant $\xi$.

The total Hamiltonian, $H_0$, of the LH1-RC system without light is therefore
\begin{eqnarray}
\label{eq:h0}
    H_0&&=H_d+H_a+H_{da},\nonumber\\
    =&&\sum_{j=1}^N[(\omega_d-i\kappa)e_j^\dagger e_j+g(e^\dagger_je_{j+1}+h.c.)+\xi(e_j^\dagger a+h.c.)]\nonumber\\
    &&+(\omega_a-i\Gamma)a^\dagger a.
\end{eqnarray}
We consider double excitation transfer from the ring to the 2-LS acceptor. The wave function of the system at time $t$ is
\begin{eqnarray}
\label{eqn:wavefunc}
    |\psi_0(t)\rangle=&&\sum^N_{i=1}\sum^N_{j>i}u_{0ij}(t)|0..1_i..1_j..0;0\rangle\nonumber\\
    &&+\sum^N_{i=1}v_{0i}(t)|0..2_i..0;0\rangle\nonumber\\
    &&+\sum^N_{i=1}w_{0i}(t)|0..1_i..0;1\rangle.
\end{eqnarray}
where for example $|0..1_1..1_2..0;0\rangle=|110..0;0\rangle$ describes the basis state with single excitations on ring sites $1$ and $2$ with corresponding amplitude $u_{012}(t)$; $|0..2_1..0;0\rangle=|20..0;0\rangle$ is the basis state with a double excitation on ring site $1$ with corresponding amplitude $v_{01}(t)$; and $|0..1_1..0;1\rangle=|10..0;1\rangle$ has a single excitation on ring site $1$ as well as on the acceptor with amplitude $w_{01}(t)$.

\subsubsection{\label{sec:lrs1}Collective LH1-RC Hamiltonian and Transfer Efficiency}
The Hamiltonian $H_{0}$ in Eq.~(\ref{eq:h0}) can be rewritten in terms of the $k-$space collective operators $\Tilde{\textbf{e}}_{0}$, $\Tilde{\textbf{e}}^\dagger_{0}$ as follows~\cite{collectop}
\begin{equation}
\label{eq:collectham}
    \Tilde{H}_{0}=(\omega_0+2g)\Tilde{\textbf{e}}^\dagger_0\Tilde{\textbf{e}}_0+\omega_1 a^\dagger a+\sqrt{N}\xi(\Tilde{\textbf{e}}^\dagger_0a+h.c.),
\end{equation}
since only $k=0$ mode couples to the acceptor (see Appendix~\ref{appen}). We define frequencies $\omega_0=\omega_d-i\kappa$ and $\omega_1=\omega_a-i\Gamma$. $\Tilde{\textbf{e}}_{0}$ and $\Tilde{\textbf{e}}^\dagger_{0}$ are defined as follows
\begin{subequations}
\label{eq:kspace}
\begin{equation}
   \Tilde{\textbf{e}}_{0}=\frac{1}{\sqrt{N}}\sum_{j=1}^Ne_j,
\end{equation}
\begin{equation}
    \Tilde{\textbf{e}}^\dagger_{0}=\frac{1}{\sqrt{N}}\sum_{j=1}^Ne^\dagger_j.
\end{equation}
\end{subequations}
where $e_j$ and $e^\dagger_j$ are defined in Eq.~(\ref{eq:ejdef}). The matrix form of $\Tilde{H}_{0}$ in Eq.~(\ref{eq:collectham}) in the double excitation subspace is given as
\begin{equation}
    \Tilde{H}_{0}=\begin{pmatrix}
        2\omega_0+4g&\sqrt{2N}\xi\\\sqrt{2N}\xi&\omega_0+\omega_1+2g
    \end{pmatrix},
\end{equation}
where the double excitation basis vectors $|\textbf{2}_N;0\rangle$ and $|\textbf{1}_N;1\rangle$ are given as
\begin{subequations}
\begin{eqnarray}
\label{eq:highnbasis}
    |\textbf{2}_N;0\rangle=&&\frac{1}{N}\biggl(\sqrt{2}\sum^N_{i=1}\sum^N_{j>i}|0..1_i..1_j..0;0\rangle\nonumber\\
    &&+\sum^N_{i=1}|0..2_i..0;0\rangle\biggl),
\end{eqnarray}
\begin{equation}
\label{eq:highnbasis2}
    |\textbf{1}_N;1\rangle=\frac{1}{\sqrt{N}}\sum^N_{i=1}|0..1_i..0;1\rangle.
\end{equation}  
\end{subequations}
The eigenvalues of $\Tilde{H}_{0}$ for $\Gamma=\kappa$ are given as
\begin{eqnarray}
    \epsilon^{(1,2)}_{0}=\frac{1}{2}\left(6g+3\omega_0+\omega_1\pm q_0\right),
\end{eqnarray}
where $q_0=\sqrt{8\xi^2N+(2g+\Delta)^2}$ and $\Delta=\omega_d-\omega_a$.

We consider an initial state at $t=0$, i.e. $|\Tilde{\Psi}_{0}(0)\rangle=|\textbf{2}_N;0\rangle$, with two excitations on the ring. The state at time $t$, $|\Tilde{\Psi}_{0}(t)\rangle$, is found by solving the Schr\"odinger equation $\frac{d|\Tilde{\Psi}_{0}(t)\rangle}{dt}=-i\Tilde{H}_0|\Tilde{\Psi}_{0}(t)\rangle$. We obtain the state $|\Tilde{\Psi}_{0}(t)\rangle$ as
\begin{eqnarray}
    |\Tilde{\Psi}&&_{0}(t)\rangle=\frac{1}{r_2}\exp\left(-\tfrac{i}{2}\left(\epsilon^{(1)}_{0}+\epsilon^{(2)}_{0}\right)t\right)\biggl(q_0\cos\left(\frac{q_0}{2}t\right)\nonumber\\
    &&-i\left(2g+\Delta\right)\sin\left(\frac{q_0}{2}t\right)\biggl)|\textbf{2}_N;0\rangle-\Bigg(\frac{i2\sqrt{2N}\xi}{q_0}\times\nonumber\\
    &&\exp\left(-\tfrac{i}{2}\left(\epsilon^{(1)}_{0}+\epsilon^{(2)}_{0}\right)t\right)\sin\left(\frac{q_0}{2}t\right)\Bigg)|\textbf{1}_N;1\rangle.
\end{eqnarray}
The probability that the acceptor is excited, $\text{P}^{(1)}(t)$, is
\begin{equation}
\label{eq:p10}
    \text{P}^{(1)}(t)=\frac{4\xi^2}{4\xi^2+\frac{(2g+\Delta)^2}{2N}}e^{-4\kappa t}\sin^2\left(\frac{q_0}{2}t\right).
\end{equation}
$\text{P}^{(1)}(t)$ can be used to find the transfer efficiency, $\eta^{(1)}_0$, as~\cite{effic1}
\begin{equation}
\label{eq:eff0}
    \eta^{(1)}_0=\int_0^\infty2\Gamma \textrm{P}^{(1)}(t)\textrm{d}t,
\end{equation}
which gives
\begin{equation}
\label{eq:eta0n}
    \eta_{0}^{(1)}=\frac{2\xi^2N}{8\xi^2N+(2g+\Delta)^2+16\kappa^2}.
\end{equation}

\subsubsection{\label{sec:indel}Initial Delocalized Double Excitations}
For simplicity, we consider the case when $N=3$, with the following basis vectors $|D_0\rangle$, $|L_0\rangle$, and $|A_0\rangle$, given as
\begin{subequations}
\label{eq:whole}
\begin{equation}
\label{eq:whole1}
    |D_0\rangle=\frac{1}{\sqrt{3}}\left(|110;0\rangle+|101;0\rangle+|011;0\rangle\right),
\end{equation}
\begin{equation}
\label{eq:whole2}
    |L_0\rangle=\frac{1}{\sqrt{3}}\left(|200;0\rangle+|020;0\rangle+|002;0\rangle\right),
\end{equation}
\begin{equation}
\label{eq:whole3}
    |A_0\rangle=\frac{1}{\sqrt{3}}\left(|100;1\rangle+|010;1\rangle+|001;1\rangle\right).
\end{equation}
\end{subequations}
$|D_0\rangle$ ($|L_0\rangle$) describes delocalized (localized) double excitations on the ring and $|A_0\rangle$ describes double excitations including an excited acceptor. Using the Hamiltonian in Eq.~(\ref{eq:h0}) we obtain the matrix form of $H_0$ in this basis, $H'_{0}$, as
\begin{equation}
\label{eq:hp0}
    H'_{0}=\begin{pmatrix}
       2\omega_0+2g&2\sqrt{2}g&2\xi\\2\sqrt{2}g&2\omega_0&\sqrt{2}\xi\\2\xi&\sqrt{2}\xi&\omega_0+\omega_1+2g
    \end{pmatrix}.
\end{equation}
The eigenvalues of $H'_{0}$ for $\Gamma=\kappa$, $\epsilon_{0}'{}^{(i)}$ ($i=1,2,3$), are found to be
\begin{subequations}
\begin{equation}
    \epsilon_{0}'{}^{(1)}=2(\omega_0-g),
\end{equation}
\begin{equation}
    \epsilon_{0}'{}^{(2)}=\frac{1}{2}(6g+3\omega_0+\omega_1-r_0),
\end{equation}
\begin{equation}
    \epsilon_{0}'{}^{(3)}=\frac{1}{2}(6g+3\omega_0+\omega_1+r_0),
\end{equation}
\end{subequations}
where $r_0=\sqrt{24\xi^2+(2g+\Delta^2)}$. These have corresponding eigenvectors
\begin{subequations}
\begin{equation}
    |\epsilon_{0}'{}^{(1)}\rangle=\frac{|D_0\rangle-\sqrt{2}|L_0\rangle}{\sqrt{3}},
\end{equation}
\begin{eqnarray}
    |\epsilon_{0}'{}^{(2)}\rangle=\frac{1}{\sqrt{3}B_{-}}\biggl(&&\left(2g+\Delta-r_0\right)\left(\sqrt{2}|D_0\rangle+|L_0\rangle\right)\nonumber\\
    &&+6\sqrt{2}\xi|A_0\rangle\biggl),
\end{eqnarray}
\begin{eqnarray}
    |\epsilon_{0}'{}^{(2)}\rangle=\frac{1}{\sqrt{3}B_{+}}\biggl(&&\left(2g+\Delta+r_0\right)\left(\sqrt{2}|D_0\rangle+|L_0\rangle\right)\nonumber\\
    &&+6\sqrt{2}\xi|A_0\rangle\biggl),
\end{eqnarray}
\end{subequations}
with $B_{\pm}=\sqrt{24\xi^2+(2g+\Delta\pm r_0)^2}$ as normalization factors. The wave function at time $t$, $|\Psi'_{0}(t)\rangle$, is given
\begin{equation}
    |\Psi'_{0}(t)\rangle=u'_{0}(t)|D_{0}\rangle+v'_{0}(t)|L_0\rangle+w'_{0}(t)|A_0\rangle.
\end{equation}
where $u'_{0}(t)$, $v'_{0}(t)$, and $w'_{0}(t)$ are the amplitudes corresponding to states $|D_{0}\rangle$, $|L_{0}\rangle$, and $|A_{0}\rangle$, respectively.
For an initial delocalized state, $|\Psi'_{0D}(0)\rangle=|D_0\rangle$, the amplitudes for the state at time $t$, $u_{0D}'(t)$, $v_{0D}'(t)$ and $w_{0D}'(t)$, are given as
\begin{subequations}
\label{eq:psi0s}
\begin{eqnarray}
    u_{0D}'(t)=&&\frac{1}{6r_0}\Biggl(4\exp\left(-\frac{i}{2}\left(\epsilon'{}_{0}^{(2)}+\epsilon'{}_{0}^{(3)}\right)t\right)\nonumber\\
    &&\times\left(r_0\cos\left(\tfrac{r_0}{2}t\right)-i(2d+\Delta)\sin\left(\tfrac{r_0}{2}t\right)\right)\nonumber\\
    &&+2r_0\exp\left(-i\epsilon'{}_{0}^{(1)}t\right)\Biggl),
\end{eqnarray}
\begin{eqnarray}
    v'_{0D}(t)=&&\frac{1}{6r_0}\Biggl(2\sqrt{2}\exp\left(-\frac{i}{2}\left(\epsilon'{}_{0}^{(2)}+\epsilon'{}_{0}^{(3)}\right)t\right)\nonumber\\
    &&\times\left(r_0\cos\left(\tfrac{r_0}{2}t\right)-i(2d+\Delta)\sin\left(\tfrac{r_0}{2}t\right)\right)\nonumber\\
    &&-2\sqrt{2}r_0\exp\left(-i\epsilon'{}_{0}^{(1)}t\right)\Biggl),
\end{eqnarray}
\begin{align}
    w'_{0D}(t)=-\frac{4\xi i}{r_0}\exp\left(-\frac{i}{2}\left(\epsilon'{}_{0}^{(2)}+\epsilon'{}_{0}^{(3)}\right)t\right)\sin\left(\frac{r_0}{2}t\right).
\end{align}
\end{subequations}
The probability for an excitation on the acceptor, $\text{P}^{(1)}_{0D}(t)$, is found to be
\begin{eqnarray}
\label{eq:p10d}
    \text{P}^{(1)}_{0D}(t)=&&\frac{16\xi^2}{24\xi^2+(2g+\Delta)^2}e^{-4\kappa t}\sin^2\left(\frac{r_0}{2}t\right).
\end{eqnarray}
$\text{P}^{(1)}_{0D}(t)$ then gives a transfer efficiency $\eta^{(1)}_{0D}$ as
\begin{equation}
    \eta^{(1)}_{0D}=\frac{4\xi^2}{24\xi^2+(2g+\Delta)^2+16\kappa^2}.
\end{equation}

\subsubsection{\label{sec:inloc}Initial Localized Double Excitations}
In this case, the localized initial state, $|\Psi'_{0L}(0)\rangle=|L_0\rangle$, given in Eq.~(\ref{eq:whole}). The amplitudes $u'_{0L}(t)$, $v'_{0L}(t)$ and $w'_{0L}(t)$, for the state at time $t$ are as follows;
\begin{subequations}
\begin{eqnarray}
    u'_{0L}(t)=&&\frac{1}{6r_0}\Biggl(2\sqrt{2}\exp\left(-\frac{i}{2}\left(\epsilon'_{0}{}^{(2)}+\epsilon'_{0}{}^{(3)}\right)t\right)\nonumber\\
    &&\times\left(r_0\cos\left(\tfrac{r_0}{2}t\right)-i(2d+\Delta)\sin\left(\tfrac{r_0}{2}t\right)\right)\nonumber\\
    &&-2\sqrt{2}r_0\exp\left(-i\epsilon'_{0}{}^{(1)}t\right)\Biggl),
\end{eqnarray}
\begin{eqnarray}
    v'_{0L}(t)=&&\frac{1}{6r_0}\Biggl(2\exp\left(-\frac{i}{2}\left(\epsilon'_{0}{}^{(2)}+\epsilon'_{0}{}^{(3)}\right)t\right)\nonumber\\
    &&\left(r_0\cos\left(\tfrac{r_0}{2}t\right)-i(2d+\Delta)\sin\left(\tfrac{r_0}{2}t\right)\right)\nonumber\\
    &&+4r_0\exp\left(-i\epsilon'_{0}{}^{(1)}t\right)\Biggl),
\end{eqnarray}
\begin{align}
    w'_{0L}(t)=-\frac{2\sqrt{2}\xi i}{r_0}\exp\left(-\frac{i}{2}\left(\epsilon'_{0}{}^{(2)}+\epsilon'_{0}{}^{(3)}\right)t\right)\sin\left(\frac{r_0}{2}t\right).
\end{align}
\end{subequations}
The probability for an excited acceptor $\text{P}^{(1)}_{0L}(t)$, is
\begin{eqnarray}
\label{eq:p10l}
    \text{P}^{(1)}_{0L}(t)=&&\frac{8\xi^2}{24\xi^2+(2g+\Delta)^2}e^{-4\kappa t}\sin^2\left(\frac{r_0}{2}t\right).
\end{eqnarray}
The transfer efficiency in this case, $\eta^{(1)}_{0L}$, is calculated to be
\begin{equation}
    \eta^{(1)}_{0L}=\frac{2\xi^2}{24\xi^2+(2g+\Delta)^2+16\kappa^2}.
\end{equation}

\subsection{\label{sec:3lsacc}LH1 Ring Coupled to a 3-LS Acceptor}
The acceptor site is a 3-LS in the ladder configuration with allowed transition frequency $\omega_a$ and spontaneous decay $\Gamma$ to the RC (see Fig.~\ref{fig:pic}(c)). The acceptor Hamiltonian $H_b$, is given as
\begin{eqnarray}
    H_b=(\omega_a-i\Gamma)b^\dagger b.
\end{eqnarray}
$b^\dagger$ ($b$) is the 3-LS creation (annihilation) operator acting on the acceptor, and defined as
\begin{eqnarray}
    b^\dagger=|1\rangle\langle0|+\sqrt{2}|2\rangle\langle1|,\nonumber\\
    b=|0\rangle\langle1|+\sqrt{2}|1\rangle\langle2|.
\end{eqnarray}
The interaction Hamiltonian describing the coupling between the acceptor and the ring, $H_{db}$, is given as
\begin{eqnarray}
    H_{db}=&&\zeta\sum_{j=1}^N(e_j^\dagger b+e_j b^\dagger),
\end{eqnarray}
where $\zeta$ is the donor-acceptor coupling constant. The total Hamiltonian, $H_{1}$, describing the LH1-RC system is 
\begin{eqnarray}
\label{eq:hamh1}
    &&H_{1}=H_d+H_b+H_{db}\\
    &&=\sum_{j=1}^N[(\omega_d-i\kappa)e_j^\dagger e_j+g(e^\dagger_je_{j+1}+h.c.)+\zeta(e_j^\dagger a+h.c.)]\nonumber\\
    &&\quad+(\omega_a-i\Gamma)a^\dagger a.
\end{eqnarray}
The generalized wave function at time $t$, $|\Psi_1(t)\rangle$, in the double excitation subspace is given as
\begin{eqnarray}
    |\Psi_1(t)\rangle=&&\sum^N_{i=1}\sum^N_{j>i}u_{1ij}(t)|0..1_i..1_j..0;0\rangle\nonumber\\
    &&+\sum^N_{i=1}v_{1i}(t)|0..2_i..0;0\rangle\\
    &&+\sum^N_{i=1}w_{1i}(t)|0..1_i..0;1\rangle+x_{1}(t)|0..0;2\rangle.\nonumber
\end{eqnarray}
where $u_{1ij}(t)$, $v_{1i}(t)$, $w_{1i}(t)$, and $x_{1}(t)$ are the amplitudes for the respective basis states $|0..1_i..1_j..0;0\rangle$, $|0..2_i..0;0\rangle$, $|0..1_i..0;1\rangle$, and $|0..0;2\rangle$.

\subsubsection{\label{sec:lrs2}Collective LH1-RC Hamiltonian and Transfer Efficiency}
For the $k-$space operators defined in Sec.~\ref{sec:lrs1}, the 3-LS acceptor couples to the $k=0$ mode defined in Eq.~(\ref{eq:kspace}). In terms of these collective operators, the Hamiltonian $H_1$ becomes
\begin{equation}
    \Tilde{H}_{1}=(\omega_0+2g)\Tilde{\textbf{e}}^\dagger_0\Tilde{\textbf{e}}_0+\omega_1 b^\dagger b+\sqrt{N}\zeta(\Tilde{\textbf{e}}^\dagger_0b+h.c.).
\end{equation}
The matrix form of $\Tilde{H}_1$ is given as
\begin{equation}
    \Tilde{H}_{1}=\begin{pmatrix}
        2\omega_0+4g&\sqrt{2N}\zeta&0\\\sqrt{2N}\zeta&\omega_0+\omega_1+2g&\sqrt{2N}\zeta\\0&\sqrt{2N}\zeta&2\omega_1
    \end{pmatrix},
\end{equation}
which acts on the basis states $|\textbf{2}_N;0\rangle, |\textbf{1}_N;1\rangle$ and $|\textbf{0}_N;2\rangle$, where $|\textbf{0}_N;2\rangle$ is defined by
\begin{eqnarray}
\label{eq:0n2}
    |\textbf{0}_N;2\rangle=|0...0;2\rangle,
\end{eqnarray}
and $|\textbf{2}_N;0\rangle$ and $|\textbf{1}_N;1\rangle$ are defined in Eq.~(\ref{eq:highnbasis}) and Eq.~(\ref{eq:highnbasis2}), respectively. The eigenvalues of $\Tilde{H}_1$ with $\Gamma=\kappa$ are
\begin{align}
    \epsilon^{(1)}_{1}=\omega_0+\omega_1&+2g-q_1,~\epsilon^{(2)}_{1}=\omega_0+\omega_1+2g,\\
    &\epsilon^{(3)}_{1}=\omega_0+\omega_1+2g+q_1,
\end{align}
with $q_1=\sqrt{4\zeta^2N+(2g+\Delta)^2}$. Starting with an initial state $|\Tilde{\Psi}_{1}(0)\rangle=|\textbf{2}_N;0\rangle$, the state at time $t$, $|\Tilde{\Psi}_{1}(t)\rangle$, is given as
\begin{align}
    |\Tilde{\Psi}_{1}(t)\rangle&=\frac{1}{q_1^2}\exp\left(-i\epsilon_1^{(2)}t\right)\biggl(2\zeta^2N\Bigl(1+\nonumber\cos\left(q_1 t\right)\Bigl)\\
    &+(2g+\Delta)^2(\cos\left(q_1 t\right)-iq_1\sin\left(q_1 t\right))\biggl)|\textbf{2}_N;0\rangle\nonumber\\
    &+\frac{\sqrt{2N}\zeta}{q_1^2}\exp\left(-i\epsilon_1^{(2)}t\right)\Bigl(\left(2g+\Delta\right)\left(\cos\left(q_1 t\right)-1\right)\nonumber\\
    &-q_1i\sin\left(q_1t\right)\Bigl)|\textbf{1}_N;1\rangle\nonumber\\
    &-\frac{2\zeta^2N}{q^2_1}\exp\left(-i\epsilon_1^{(2)}t\right)\left(1-\cos\left(q_1 t\right)\right)|\textbf{0}_N;2\rangle.
\end{align}
The probabilities that the acceptor is singly-excited, $\text{P}^{(1)}_1$, and doubly-excited, $\text{P}^{(2)}_1$, are given as
\begin{subequations}
\begin{align}
\label{eq:ps1n}
    \text{P}^{(1)}_1(t)=&\frac{2\zeta^2N}{q_1^4}\Bigl(\left(2g+\Delta\right)^2\left(1-\cos\left(q_1 t\right)\right)^2\nonumber\\
    &+q_1^2\sin(q_1 t)^2\Bigl)e^{-4\kappa t},
\end{align}
\begin{align}
\label{eq:p1n}
    \text{P}^{(2)}_{1}(t)=\frac{4\zeta^4N^2}{(4N\zeta^2+(2g+\Delta)^2)^2}(1-\cos(q_1 t))^2e^{-4\kappa t}.
\end{align}
\end{subequations}
The double excitation transfer efficiency $\eta^{(2)}_{1}$ is defined as
\begin{align}
\label{eq:eta1def}
    \eta^{(2)}_{1}=\int_0^\infty4\Gamma\text{P}^{(2)}_{1}(t)\text{d}t.
\end{align}
Substituting Eq.~(\ref{eq:p1n}) in Eq.~(\ref{eq:eta1def}) gives $\eta^{(2)}_{1}$ as
\begin{equation}
\label{eq:eta1}
    \resizebox{.9\hsize}{!}{$\eta^{(2)}_{1}=\frac{6\zeta^4N^2}{(4\zeta^2N+16\kappa^2+(2g+\Delta)^2)(4\zeta^2N+4\kappa^2+(2g+\Delta)^2)}$}.
\end{equation}
The probability of transferring a single excitation, $\eta^{(1)}_{1}$, defined in Eq.~(\ref{eq:eff0}) is 
\begin{align}
\label{eq:etat}
    \resizebox{.9\hsize}{!}{$\eta^{(1)}_{1}=\frac{2\zeta^2N(\zeta^2N+4\kappa^2+(2g+\Delta)^2)}{(4\zeta^2N+16\kappa^2+(2g+\Delta)^2)(4\zeta^2N+4\kappa^2+(2g+\Delta)^2)}$}.
\end{align}

\subsubsection{\label{sec:dsc}Initial Delocalized Double Excitations}
Using the Hamiltonian in Eq.~(\ref{eq:hamh1}), with the basis states in Eqs.~(\ref{eq:whole1})-(\ref{eq:whole3}) (which also apply to a 2-LS acceptor), in addition to $|A_1\rangle=|\textbf{0}_3;2\rangle$ from Eq.~(\ref{eq:0n2}), the matrix form of $H_1$, which is denoted $H'_{1}$, is
\begin{equation}
    H'_{1}=\begin{pmatrix}
        2\omega_0+2g&2\sqrt{2}g&2\zeta&0\\2\sqrt{2}g&2\omega_0&\sqrt{2}\zeta&0\\2\zeta&\sqrt{2}\zeta&\omega_0+\omega_1+2g&\sqrt{6}\zeta\\0&0&\sqrt{6}\zeta&2w_1
    \end{pmatrix},
\end{equation}
with eigenvalues $\epsilon_1'{}^{(i)} (i=1,..,4)$, for $\Gamma=\kappa$ given as
\begin{align}
    \epsilon_{1}'{}^{(1)}=2(\omega_0-g),~\epsilon_{1}'{}^{(2)}=2g+\omega_0+\omega_1-r_1,\nonumber\\
    \epsilon_{1}'{}^{(3)}=2g+\omega_0+\omega_1,~\epsilon_{1}'{}^{(4)}=2g+\omega_0+\omega_1+r_1,
\end{align}
where $r_1=\sqrt{12\zeta^2+(2g+\Delta)^2}$. The wave function at time $t$, $|\Psi'_{1}(t)\rangle$, is
\begin{eqnarray}
    |\Psi_{1}(t)\rangle=&&u'_{1}(t)|D_{0}\rangle+v'_{1}(t)|L_0\rangle+w'_{1}(t)|A_0\rangle\nonumber\\
    &&+x'_1(t)|A_1\rangle,
\end{eqnarray}
with amplitudes $u'_{1}(t)$, $v'_{1}(t)$, $w'_{1}(t)$, and $x'_{1}(t)$ corresponding to the states $|D_{0}\rangle$, $|L_{0}\rangle$, $|A_{0}\rangle$, $|A_{1}\rangle$, respectively. For an initial delocalized state, $|\Psi'_{1D}(0)\rangle=|D_0\rangle$, the amplitudes at time $t$ are given as
\begin{subequations}
\begin{align}
    &u'_{1D}(t)=\frac{1}{3}\exp\left(-i\epsilon'_1{}^{(1)}t\right)+\frac{1}{3r_1^2}\exp\left(-i\epsilon'_1{}^{(3)}t\right)\Bigl(12\zeta^2\nonumber\\
    &+\left((2g+\Delta)^2+r_1^2\right)\cos\left(r_1t\right)-2ir_1(2g+\Delta)\sin\left(r_1\right)\Bigl),
\end{align}
\begin{align}
    &v'_{1D}(t)=\frac{-\sqrt{2}}{3}\exp\left(-i\epsilon'_1{}^{(1)}t\right)+\frac{\sqrt{2}}{3r_1^2}\exp\left(-i\epsilon'_1{}^{(3)}t\right)\Bigl(6\zeta^2\nonumber\\
    &+\left((2g+\Delta)^2+6\zeta^2\right)\cos\left(r_1t\right)-ir_1(2g+\Delta)\sin\left(r_1\right)\Bigl),
\end{align}
\begin{align}
    w'_{1D}(t)=&\frac{2\zeta}{r_1^2}\exp\left(-i\epsilon'_1{}^{(3)}t\right)\nonumber\\
    &\Bigl(\left(2g+\Delta\right)(\cos\left(r_1t\right)-1)-ir_1\sin\left(r_1t\right)\Bigl),
\end{align}
\begin{eqnarray}
    x'_{1D}(t)=\frac{2\sqrt{6}\zeta^2}{r_1^2}\exp(-i\epsilon'_1{}^{(3)}t)(\cos(r_1t)-1).
\end{eqnarray}
\end{subequations}
The probabilities that the acceptor is singly-excited, $\text{P}^{(1)}_{1D}(t)$, and doubly-excited, $\text{P}^{(2)}_{1D}(t)$, are given as
\begin{subequations}
\label{eq:p1}
\begin{align}
\label{eq:p1d1}
    \text{P}^{(1)}_{1D}(t)=\frac{4\zeta^2}{\left(12\zeta^2+(2g+\Delta)^2\right)^2}e^{-4\kappa t}\nonumber\\
    \times\left(\left(2g+\Delta\right)^2(\cos(r_1t)-1)^2+r_1^2\sin\left(r_1t\right)\right),
\end{align}
\begin{eqnarray}
\label{eq:p1d2}
    \text{P}^{(2)}_{1D}(t)=&&\frac{24\zeta^4(\cos(r_1t)-1)^2}{(12\zeta^2+(2g+\Delta)^2)^2}e^{-4\kappa t}.
\end{eqnarray} 
\end{subequations}
The single-excitation transfer efficiency, $\eta^{(1)}_{1D}$, and double-excitation transfer efficiency, $\eta^{(2)}_{1D}$, are given as
\begin{subequations}
\begin{align}
    \resizebox{.9\hsize}{!}{$\eta^{(1)}_{1D}=\frac{4\zeta^2(3\zeta^2+4\kappa^2+(2g+\Delta)^2)}{(12\zeta^2+16\kappa^2+(2g+\Delta)^2)(12\zeta^2+4\kappa^2+(2g+\Delta)^2)}$},
\end{align}
\begin{align}
    \resizebox{.9\hsize}{!}{$\eta^{(2)}_{1D}=\frac{36\zeta^4}{(12\zeta^2+16\kappa^2+(2g+\Delta)^2)(12\zeta^2+4\kappa^2+(2g+\Delta)^2)}$}.
\end{align}
\end{subequations}

\subsubsection{Initial Localized Double Excitations}
We now consider an initial localized state $|\Psi'_{1L}(0)\rangle=|L_0\rangle$, which results in the following amplitudes in the wave function at time $t$:
\begin{subequations}
\begin{align}
    &u'_{1L}(t)=-\frac{\sqrt{2}}{3}\exp\left(-i\epsilon'_1{}^{(1)}t\right)+\frac{\sqrt{2}}{3r_1^2}\exp\left(-i\epsilon'_1{}^{(3)}t\right)\Bigl(6\zeta^2\nonumber\\
    &+\left((2g+\Delta)^2+6\zeta^2\right)\cos\left(r_1t\right)-ir_1(2g+\Delta)\sin\left(r_1\right)\Bigl),
\end{align}
\begin{align}
    &v'_{1L}(t)=\frac{2}{3}\exp\left(-i\epsilon'_1{}^{(1)}t\right)+\frac{1}{3r_1^2}\exp\left(-i\epsilon'_1{}^{(3)}t\right)\Bigl(6\xi^2\nonumber\\
    &+\left((2g+\Delta)^2+6\zeta^2\right)\cos\left(r_1t\right)-ir_1(2g+\Delta)\sin\left(r_1\right)\Bigl),
\end{align}
\begin{align}
    w'_{1L}(t)=&\frac{\sqrt{2}\zeta}{r_1^2}\exp\left(-i\epsilon'_1{}^{(3)}t\right)\nonumber\\
    &\Bigl(\left(2g+\Delta\right)(\cos\left(r_1t\right)-1)-ir_1\sin\left(r_1t\right)\Bigl),
\end{align}
\begin{align}
    x'_{1L}(t)=\frac{2\sqrt{3}\zeta^2}{r_1^2}\exp(-i\epsilon'_1{}^{(3)}(\cos(r_1t)-1),
\end{align}
\end{subequations}
The probabilities for a singly-excited and doubly-excited acceptor, $\text{P}^{(1)}_{1L}(t)$ and $\text{P}^{(2)}_{1L}(t)$, respectively, are given as
\begin{subequations}
\label{eq:probphot}
\begin{eqnarray}
\label{eq:p1l1}
    \text{P}^{(1)}_{1L}(t)=\frac{2\zeta^2}{\left(12\zeta^2+(2g+\Delta)^2\right)^2}e^{-4\kappa t}\nonumber\\
    \left(\left(2g+\Delta\right)^2(\cos(r_1t)-1)^2+r_1^2\sin\left(r_1t\right)\right),
\end{eqnarray}
\begin{eqnarray}
\label{eq:p1l2}
    \text{P}^{(2)}_{1L}(t)=&&\frac{12\zeta^4(\cos(r_1t)-1)^2}{(12\zeta^2+(2g+\Delta)^2)^2}e^{-4\kappa t}.
\end{eqnarray} 
\end{subequations}
The single-excitation and double-excitation transfer efficiencies, $\eta^{(1)}_{1L}$ and $\eta^{(2)}_{1L}$, respectively, are
\begin{subequations}
\begin{align}
    \resizebox{.9\hsize}{!}{$\eta^{(1)}_{1L}=\frac{2\zeta^2(3\zeta^2+4\kappa^2+(2g+\Delta)^2)}{(12\zeta^2+16\kappa^2+(2g+\Delta)^2)(12\zeta^2+4\kappa^2+(2g+\Delta)^2)}$},
\end{align}
\begin{align}
    \resizebox{.9\hsize}{!}{$\eta^{(2)}_{1L}=\frac{18\zeta^4}{(12\zeta^2+16\kappa^2+(2g+\Delta)^2)(12\zeta^2+4\kappa^2+(2g+\Delta)^2)}$}.
\end{align}
\end{subequations}

\subsection{\label{sec:photons}LH1 Ring Coupled to Two Photons}
The Hamiltonian which described two incident photons, $H_p$
\begin{equation}
\label{eq:hpho}
    H_p=\omega_{k_{1}}c_{k_{1}}^\dagger c_{k_{1}}+\omega_{k_{2}} c_{k_{2}}^\dagger c_{k_{2}},
\end{equation}
where the photons frequencies are $\omega_{k_{1}}$ and $\omega_{k_{2}}$, respectively. The annihilation (creation) operators, $c_{k_{i}}$ ($c_{k_{i}}^\dagger$), are defined as $|1_{k_i}\rangle\langle0_{k_i}|$ ($|0_{k_i}\rangle\langle1_{k_i}|$), $i=1,2$. $|0_{k_i}\rangle$ and $|1_{k_i}\rangle$ are the vacuum and one photon states corresponding to the mode $k_i$, respectively. Each photon is coupled to the ring with coupling constant $J$ which gives the photon pair interaction Hamiltonian $H_{dp}$ as
\begin{equation}
    H_{dp}=J\sum_{j=1}^N\left[(c_{k_{1}}^\dagger+c_{k_{2}}^\dagger)e_j+h.c.\right].
\end{equation}
The interaction Hamiltonian, $H_{dp}$, can be written in terms of the collective ring operator $\Tilde{\textbf{e}}_0$, as
\begin{eqnarray}
\label{eq:hringphoton}
    \Tilde{H}_{dp}=\sqrt{N}J((c_{k_{1}}^\dagger+c_{k_{2}}^\dagger)\Tilde{\textbf{e}}_0+h.c.).
\end{eqnarray}

\subsubsection{\label{sec:photon2ls}LH1 Ring Coupled to a 2-LS Acceptor with an Incident Photon Pair}
Combining the Hamiltonians in Eq.~(\ref{eq:hpho}) and Eq.~(\ref{eq:hringphoton}) with $\Tilde{H}_0$ in Eq.~(\ref{eq:collectham}), we obtain the total Hamiltonian $\Tilde{H}_2$, as
\begin{align}
    &\Tilde{H}_{2}=H_0+H_p+\Tilde{H}_{dp},\nonumber\\
    &=(\omega_0+2g)\Tilde{\textbf{e}}^\dagger_0\Tilde{\textbf{e}}_0+\omega_1 a^\dagger a+\sqrt{N}\xi(\Tilde{\textbf{e}}^\dagger_0a+h.c.)\nonumber\\
    &+\omega_{k_{1}}c_{k_{1}}^\dagger c_{k_{1}}+\omega_{k_{2}} c_{k_{2}}^\dagger c_{k_{2}}+\sqrt{N}J((c_{k_{1}}^\dagger+c_{k_{2}}^\dagger)\Tilde{\textbf{e}}_0+h.c.).
\end{align}
In matrix form, $\Tilde{H}_{2}$ becomes
\begin{widetext}
\begin{align}
    \Tilde{H}_{2}=\begin{pmatrix}
        \omega_{k_{1}}+\omega_{k_{2}}&\sqrt{N}J&\sqrt{N}J&0&0&0&0\\\sqrt{N}J&\omega_{k_{1}}+\omega_0+2g&0&\sqrt{2N}J&\sqrt{N}\xi&0&0\\\sqrt{N}J&0&\omega_{k_{2}}+\omega_0+2g&\sqrt{2N}J&0&\sqrt{N}\xi&0\\0&\sqrt{2N}J&\sqrt{2N}J&2\omega_0+4g&0&0&\sqrt{2N}\xi\\0&\sqrt{N}\xi&0&0&\omega_{k_{1}}+\omega_1&0&\sqrt{N}J\\0&0&\sqrt{N}\xi&0&0&\omega_{k_{2}}+\omega_1&\sqrt{N}J\\0&0&0&\sqrt{2N}\xi&\sqrt{N}J&\sqrt{N}J&\omega_0+\omega_1+2g
    \end{pmatrix},
\end{align}
\end{widetext}
which acts on the basis states
\begin{align}
\label{eq:photonbasis}
    |1_{k_1}1_{k_2};\textbf{0}_N;0\rangle,&|1_{k_1}0_{k_2};\textbf{1}_N;0\rangle,|0_{k_1}1_{k_2};\textbf{1}_N;0\rangle,\nonumber\\
    |0_{k_1}0_{k_2};\textbf{2}_N;0\rangle,&|1_{k_1}0_{k_2};\textbf{0}_N;1\rangle,|1_{k_1}0_{k_2};\textbf{0}_N;1\rangle,\nonumber\\
    &|0_{k_1}0_{k_2};\textbf{1}_N;1\rangle.
\end{align} 
The wave function at time $t$, $|\Tilde{\Psi}_2(t)\rangle$, is given as
\begin{align}
\label{eq:photonwf}
    |\Tilde{\Psi}_2(t)\rangle=&\Tilde{n}_2(t)|1_{k_1}1_{k_2};\textbf{0}_N;0\rangle+\Tilde{p}_2(t)|1_{k_1}0_{k_2};\textbf{1}_N;0\rangle\nonumber\\
    &+\Tilde{q}_2(t)|0_{k_1}1_{k_2};\textbf{1}_N;0\rangle\nonumber+\Tilde{s}_2(t)|0_{k_1}0_{k_2};\textbf{2}_N;0\rangle\nonumber\\
    &+\Tilde{u}_2(t)|1_{k_1}0_{k_2};\textbf{0}_N;1\rangle+\Tilde{v}_2(t)|0_{k_1}1_{k_2};\textbf{0}_N;1\rangle\nonumber\\
    &+\Tilde{w}_2(t)|0_{k_1}0_{k_2};\textbf{1}_N;1\rangle,
\end{align}
where $\Tilde{n}_2(t)$, $\Tilde{p}_2(t)$, $\Tilde{q}_2(t)$, $\Tilde{s}_2(t)$, $\Tilde{u}_2(t)$, $\Tilde{v}_2(t)$ and $\Tilde{w}_2(t)$ are the amplitudes corresponding to the basis states defined in Eq.~(\ref{eq:photonbasis}). The transition energies of the collective donor ($\omega_d+2g$) and acceptor ($\omega_a$) are on resonance, i.e. $\omega_d+2g=\omega_a$, and the photon frequencies are detuned by $\delta$, such that $\omega_{k_{1}}=\omega_a+\delta$ and $\omega_{k_{2}}=\omega_a-\delta$. In the perturbative regime, where $J<<\xi$, the system will evolve adiabatically and we can apply perturbation theory to calculate the eigenvalues of $\Tilde{H}_2$, $\epsilon_2^{(j)} (j=1,..,7)$, with $\Gamma=\kappa$ which are found to be
\begin{subequations}
\begin{equation}
    \epsilon_2^{(1)}=2\omega_a-\frac{2NJ^2\kappa(\delta^2+N\xi^2+\kappa^2)}{\delta^4+D^2(\kappa^2-N\xi^2)+(\kappa^2+N\xi^2)^2},
\end{equation}
\begin{align}
    \epsilon_2^{(2)}=&2\omega_a-\delta-i\kappa-\sqrt{N}\xi-\frac{NJ^2}{2(\delta+i\kappa+\sqrt{N}\xi)}\nonumber\\
    &+\frac{NJ^2(3i\kappa-3\delta-7\sqrt{N}\xi)}{2((\delta-i\kappa)^2+2(\delta-i\kappa)\sqrt{N}\xi-N\xi^2)},
\end{align}
\begin{align}
    \epsilon_2^{(3)}=&2\omega_a+\delta-i\kappa-\sqrt{N}\xi+\frac{NJ^2}{2(\delta-i\kappa-\sqrt{N}\xi)}\nonumber\\
    &+\frac{NJ^2(3i\kappa+3\delta-7\sqrt{N}\xi)}{2((\delta+i\kappa)^2-2(\delta+i\kappa)\sqrt{N}\xi-N\xi^2)},
\end{align}
\begin{align}
    \epsilon_2^{(4)}=&2\omega_a-\delta-i\kappa+\sqrt{N}\xi-\frac{NJ^2}{2(\delta+i\kappa-\sqrt{N}\xi)}\nonumber\\
    &+\frac{NJ^2(3i\kappa-3\delta+7\sqrt{N}\xi)}{2((\delta-i\kappa)^2-2(\delta-i\kappa)\sqrt{N}\xi-N\xi^2)},
\end{align}
\begin{align}
    \epsilon_2^{(5)}=&2\omega_a+\delta-i\kappa+\sqrt{N}\xi+\frac{NJ^2}{2(\delta-i\kappa+\sqrt{N}\xi)}\nonumber\\
    &+\frac{NJ^2(3i\kappa+3\delta+7\sqrt{N}\xi)}{2((\delta+i\kappa)^2+2(\delta+i\kappa)\sqrt{N}\xi-N\xi^2)},
\end{align}
\begin{align}
    \epsilon_2^{(6)}=&2\omega_a-2i\kappa-\sqrt{2N}\xi\nonumber\\
    &-\frac{(2\sqrt{2}-3)NJ^2(\sqrt{N}\xi(1+\sqrt{2})+i\kappa)}{2(\delta^2-(i\kappa+(1+\sqrt{2})\sqrt{N}\xi)^2)}\nonumber\\
   & +\frac{NJ^2(i(3+2\sqrt{2})\kappa+(1+ \sqrt{2})\sqrt{N}x)}{2(\delta^2+(i\kappa+(\sqrt{2}-1)\xi)^2)},
\end{align}
\begin{align}
    \epsilon_2^{(7)}=&2\omega_a-2i\kappa+\sqrt{2N}\xi\nonumber\\
    &+\frac{(2\sqrt{2}-3)NJ^2(\sqrt{N}\xi(1+\sqrt{2})-I\kappa)}{2(\delta^2-(i\kappa-(1+\sqrt{2})\xi)^2)}\nonumber\\
    &+\frac{NJ^2((3+2\sqrt{2})\kappa+i(1+\sqrt{2})\sqrt{N}x)}{2(\delta^2-(i\kappa+(1-\sqrt{2})\sqrt{N}\xi)^2)}.
\end{align}
\end{subequations}
Starting in the state $|\Tilde{\Psi}_2(0)\rangle=|1_{k_1}1_{k_2};\textbf{0}_N;0\rangle$, the amplitudes at time $t$ in Eq.~(\ref{eq:photonwf}) are obtained as
\begin{subequations}
\begin{align}
    &\Tilde{n}_2(t)=-\sum^7_{j=1}\frac{\exp\left(-i\epsilon^{(j)}_2t\right)}{\prod_{k\neq j}\left(\epsilon^{(j)}_2-\epsilon^{(k)}_2\right)}\nonumber\\
    &\times\left(\left(\alpha_1-\epsilon_2^{(j)}\right)-N\xi^2\right)\left(\left(\alpha_2-\epsilon_2^{(j)}\right)-N\xi^2\right)\nonumber\\
    &\times\left(\left(2\kappa+2i\omega_a-i\epsilon_2^{(j)}\right)^2+2N\xi^2\right),
\end{align}
\begin{align}
    &\Tilde{p}_2(t)=-\sqrt{N}J\sum^7_{j=1}\frac{\exp\left(-i\epsilon^{(j)}_2t\right)}{\prod_{k\neq j}\left(\epsilon^{(j)}_2-\epsilon^{(k)}_2\right)}\left(\epsilon_2^{(j)}-\alpha_1\right)\times\nonumber\\
    &\left(\left(\alpha_2-\epsilon_2^{(j)}\right)-N\xi^2\right)\left(\left(2\kappa+2i\omega_a-i\epsilon_2^{(j)}\right)^2+2N\xi^2\right),
\end{align}
\begin{align}
    &\Tilde{q}_2(t)=-\sqrt{N}J\sum^7_{j=1}\frac{\exp\left(-i\epsilon^{(j)}_2t\right)}{\prod_{k\neq j}\left(\epsilon^{(j)}_2-\epsilon^{(k)}_2\right)}\left(\epsilon_2^{(j)}-\alpha_2\right)\times\nonumber\\
    &\left(\left(\alpha_1-\epsilon_2^{(j)}\right)-N\xi^2\right)\nonumber\left(\left(2\kappa+2i\omega_a-i\epsilon_2^{(j)}\right)^2+2N\xi^2\right),
\end{align}
\begin{align}
    &\Tilde{s}_2(t)=2\sqrt{2}NJ^2\xi\sum^7_{j=1}\frac{\exp\left(-i\epsilon^{(j)}_2t\right)}{\prod_{k\neq j}\left(\epsilon^{(j)}_2-\epsilon^{(k)}_2\right)}\nonumber\\
    &\times\Bigg(\left(N\xi^2+\left(\kappa+2i\omega_a-i\epsilon^{(j)}_2\right)^2\right)\nonumber\\
    &\times\left(2\kappa^2-N\xi^2+3i\kappa\left(2\omega_a-\epsilon^{(j)}_2\right)-\left(2\omega_1-\epsilon^{(j)}_2\right)^2\right)\nonumber\\
    &+\delta^2\left(2\kappa^2+N\xi^2+3i\kappa\left(2\omega_a-\epsilon^{(j)}_2\right)-(\epsilon^{(j)}_2-2\omega_a)^2\right)\Bigg),
\end{align}
\begin{align}
\label{eq:ph2lsu}
    &\Tilde{u}_2(t)=-NJ\xi\sum^7_{j=1}\frac{\exp\left(-i\epsilon^{(j)}_2t\right)}{\prod_{k\neq j}\left(\epsilon^{(j)}_2-\epsilon^{(k)}_2\right)}\times\nonumber\\
    &\left(\left(\alpha_2-\epsilon_2^{(j)}\right)-N\xi^2\right)\left(\left(2\kappa+2i\omega_a-i\epsilon_2^{(j)}\right)^2+2N\xi^2\right),
\end{align}
\begin{align}
\label{eq:ph2lsv}
    &\Tilde{v}_2(t)=-NJ\xi\sum^7_{j=1}\frac{\exp\left(-i\epsilon^{(j)}_2t\right)}{\prod_{k\neq j}\left(\epsilon^{(j)}_2-\epsilon^{(k)}_2\right)}\times\nonumber\\
    &\left(\left(\alpha_1-\epsilon_2^{(j)}\right)-N\xi^2\right)\left(\left(2\kappa+2i\omega_a-i\epsilon_2^{(j)}\right)^2+2N\xi^2\right),
\end{align}
\begin{align}
\label{eq:ph2lsw}
    &\Tilde{w}_2(t)=-2\sqrt{N^3}J^2\xi\sum^7_{j=1}\frac{\exp\left(-i\epsilon^{(j)}_2t\right)}{\prod_{k\neq j}\left(\epsilon^{(j)}_2-\epsilon^{(k)}_2\right)}\nonumber\\
    &\times\Bigg(\delta^2\left(\epsilon_2^{(j)}-2\omega_a\right)+\left(\kappa+i\left(2\omega_a-x-\epsilon_2^{(j)}\right)\right)\nonumber\\
    &\times\left(4\kappa+6i\omega_a-3i\epsilon_2^{(j)}\right)\left(i\kappa-2\omega_a-\sqrt{N}\xi+\epsilon_2^{(j)}\right)\Bigg),
\end{align}
\end{subequations}
where $\alpha_1=2\omega_a+\delta-i\kappa$ and $\alpha_2=2\omega_a-\delta-i\kappa$. The probability that the acceptor is singly-excited, $\text{P}^{(1)}_2(t)$, is
\begin{eqnarray}
    \text{P}^{(1)}_2(t)=|\Tilde{u}_2(t)|^2+|\Tilde{v}_2(t)|^2+|\Tilde{w}_2(t)|^2.
\end{eqnarray}
The long-time behavior of $\Tilde{u}_2(t)$, $\Tilde{v}_2(t)$, and $\Tilde{w}_2(t)$ to a good approximation is determined by the term with $\exp(-i\epsilon_2^{(1)}t)$ in Eqs.~(\ref{eq:ph2lsu})-(\ref{eq:ph2lsw}). We can therefore approximate the probabilities $|\Tilde{u}_2(t)|^2$, $|\Tilde{v}_2(t)|^2$, and $|\Tilde{w}_2(t)|^2$ as
\begin{subequations}
\begin{eqnarray}
    |\Tilde{u}_2(t)|^2=|\Tilde{v}_2(t)|^2\approx\frac{N^2J^2\xi^2}{F_2}\exp\left(-\frac{E_2}{F_2}t\right),
\end{eqnarray}
\begin{align}
    |\Tilde{w}_2(t)|^2\approx\frac{16N^3J^4\xi^2\kappa^2(\kappa^2+N\xi^2)^2\exp\left(-\frac{E_2}{F_2}t\right)}{F_2^2(2\kappa^2+N\xi^2)^2},
\end{align}
\end{subequations}
where
\begin{eqnarray}
    &&E_2=4NJ^2 \kappa(\delta^2+\kappa^2+N\xi^2),\nonumber\\
    &&F_2=(\delta^4+2\delta^2(\kappa^2-N\xi^2)+(\kappa^2+N\xi^2)^2).
\end{eqnarray}
To lowest-order of $J$, i.e. $J^2$, $\text{P}^{(1)}_2(t)$ is
\begin{align}
    \text{P}^{(1)}_2(t)=&|\Tilde{u}_2(t)|^2+|\Tilde{v}_2(t)|^2\nonumber\\
    =&\frac{2N^2J^2\xi^2}{F_2}\exp\left(-\frac{E_2}{F_2}t\right).
\end{align}
This gives the single-excitation transfer efficiency $\eta^{(1)}_2$, as
\begin{eqnarray}
\label{eq:eta12}
    \eta^{(1)}_2\approx\frac{N\xi^2}{\delta^2+\kappa^2+N\xi^2}.
\end{eqnarray}

\subsubsection{LH1 Ring Coupled to a 3-LS Acceptor with an Incident Photon Pair}
Considering our acceptor is a 3-LS, the total system Hamiltonian $\Tilde{H}_3$ is given as
\begin{align}
    &\Tilde{H}_{3}=\Tilde{H}_1+H_p+\Tilde{H}_{dp},\nonumber\\
    &=(\omega_0+2g)\Tilde{\textbf{e}}^\dagger_0\Tilde{\textbf{e}}_0+\omega_1 b^\dagger b+\sqrt{N}\zeta(\Tilde{\textbf{e}}^\dagger_0b+h.c.)\nonumber\\
    &+\omega_{k_{1}}c_{k_{1}}^\dagger c_{k_{1}}+\omega_{k_{2}} c_{k_{2}}^\dagger c_{k_{2}}+\sqrt{N}J((c_{k_{1}}^\dagger+c_{k_{2}}^\dagger)\Tilde{\textbf{e}}_0+h.c.).
\end{align}
where $\Tilde{H}_1$ is defined in Eq.~(\ref{eq:hamh1}), $H_p$ in Eq.~(\ref{eq:hpho}) and $\Tilde{H}_{pd}$ in Eq.~(\ref{eq:hringphoton}). In matrix form, $\Tilde{H}_3$ becomes
\begin{widetext}
\begin{align}
\label{eq:3lspheqn}
    \Tilde{H}_{3}=\begin{pmatrix}
        \omega_{k_{1}}+\omega_{k_{2}}&\sqrt{N}J&\sqrt{N}J&0&0&0&0&0\\\sqrt{N}J&\omega_{k_{1}}+\omega_0+2g&0&\sqrt{2N}J&\sqrt{N}\zeta&0&0&0\\\sqrt{N}J&0&\omega_{k_{2}}+\omega_0+2g&\sqrt{2N}J&0&\sqrt{N}\zeta&0&0\\0&\sqrt{2N}J&\sqrt{2N}J&2\omega_0+4g&0&0&\sqrt{2N}\zeta&0\\0&\sqrt{N}\zeta&0&0&\omega_{k_{1}}+\omega_1&0&\sqrt{N}J&0\\0&0&\sqrt{N}\zeta&0&0&\omega_{k_{2}}+\omega_1&\sqrt{N}J&0\\0&0&0&\sqrt{2N}\zeta&\sqrt{N}J&\sqrt{N}J&\omega_0+\omega_1+2g&\sqrt{2N}\zeta\\0&0&0&0&0&0&\sqrt{2N}\zeta&2\omega_1
    \end{pmatrix},
\end{align}
\end{widetext}
which acts on the same basis states in Eq.~(\ref{eq:photonbasis}), in addition to the state $|0_{k_1}0_{k_2};\textbf{0}_N;2\rangle$. The wave function at time $t$, $|\Tilde{\Psi}_3(t)\rangle$ is
\begin{align}
\label{eq:photonwf3ls}
    |\Tilde{\Psi}_3(t)\rangle=&\Tilde{n}_3(t)|1_{k_1}1_{k_2};\textbf{0}_N;0\rangle+\Tilde{p}_3(t)|1_{k_1}0_{k_2};\textbf{1}_N;0\rangle\nonumber\\
    &+\Tilde{q}_3(t)|0_{k_1}1_{k_2};\textbf{1}_N;0\rangle\nonumber+\Tilde{s}_3(t)|0_{k_1}0_{k_2};\textbf{2}_N;0\rangle\nonumber\\
    &+\Tilde{u}_3(t)|1_{k_1}0_{k_2};\textbf{0}_N;1\rangle+\Tilde{v}_3(t)|0_{k_1}1_{k_2};\textbf{0}_N;1\rangle\nonumber\\
    &+\Tilde{w}_3(t)|0_{k_1}0_{k_2};\textbf{1}_N;1\rangle+\Tilde{x}_3(t)|0_{k_1}0_{k_2};\textbf{0}_N;2\rangle,
\end{align}
where $\Tilde{n}_3(t)$, $\Tilde{p}_3(t)$, $\Tilde{q}_3(t)$, $\Tilde{s}_3(t)$, $\Tilde{u}_3(t)$, $\Tilde{v}_3(t)$, $\Tilde{w}_3(t)$ and $\Tilde{x}_3(t)$ are the amplitudes corresponding to the basis states. 

Similar to the case with the 2-LS acceptor, transition energies of the collective donor ($\omega_d+2g$) and acceptor ($\omega_a$) are on resonance, i.e. $\omega_d+2g=\omega_a$, and the pair photon frequencies are detuned by $\delta$, such that $\omega_{k_{1}}=\omega_a+\delta$ and $\omega_{k_{2}}=\omega_a-\delta$. For $J<<\zeta$, we can calculate the eigenvalues of $\Tilde{H}_3$, $\epsilon_3^{(j)} (j=1,..,8)$, for $\Gamma=\kappa$ are found to be
\begin{subequations}
\begin{align}
    \epsilon^{(1)}_3=2\omega_a-\frac{2iNJ^2\kappa(\delta^2+\kappa^2+N\zeta^2}{\delta^4+2\delta^2(\kappa^2-N\zeta^2)+(\kappa^2+N\zeta^2)^2},
\end{align}
\begin{align}
    \epsilon^{(2)}_3=2\omega_a-i\kappa+\frac{2iNJ^2\kappa(\delta^2+\kappa^2+N\zeta^2}{\delta^4+2\delta^2(\kappa^2-N\zeta^2)+(\kappa^2+N\zeta^2)^2},
\end{align}
\begin{align}
    \epsilon^{(3)}_3=2(\omega_a-\sqrt{N}\zeta-i\kappa)+\frac{2NJ^2(i\kappa+\sqrt{N}\zeta)}{\delta^2-(\sqrt{N}\zeta+i\kappa)^2},
\end{align}
\begin{align}
    \epsilon^{(4)}_3=2(\omega_a+\sqrt{N}\zeta-i\kappa)+\frac{2NJ^2(i\kappa+\sqrt{N}\zeta)}{\delta^2-(\sqrt{N}\zeta-i\kappa)^2},
\end{align}
\begin{align}
    \epsilon^{(5)}_3=2\omega_a-\delta&-i\kappa-\sqrt{N}\zeta+\frac{NJ^2}{2}\Bigg(\frac{2}{-\delta+i\kappa+\sqrt{N}\zeta}\nonumber\\
    &-\frac{2(\delta+\sqrt{N}\xi)}{\delta^2+\kappa^2+N\zeta^2+2\sqrt{N}\delta\zeta}\Bigg),
\end{align}
\begin{align}
    \epsilon^{(6)}_3=2\omega_a+\delta&-i\kappa-\sqrt{N}\zeta+\frac{NJ^2}{2}\Bigg(\frac{2}{\delta+i\kappa+\sqrt{N}\zeta}\nonumber\\
    &+\frac{2(\delta-\sqrt{N}\xi)}{\delta^2+\kappa^2+N\zeta^2-2\sqrt{N}\delta\zeta}\Bigg),
\end{align}
\begin{align}
    \epsilon^{(7)}_3=2\omega_a-\delta&+i\kappa-\sqrt{N}\zeta+\frac{NJ^2}{2}\Bigg(\frac{-2}{\delta-i\kappa+\sqrt{N}\zeta}\nonumber\\
    &-\frac{2(\delta-\sqrt{N}\xi)}{\delta^2+\kappa^2+N\zeta^2-2\sqrt{N}\delta\zeta}\Bigg),
\end{align}
\begin{align}
    \epsilon^{(8)}_3=2\omega_a+\delta&-i\kappa+\sqrt{N}\zeta+\frac{NJ^2}{2}\Bigg(\frac{2}{\delta+i\kappa-\sqrt{N}\zeta}\nonumber\\
    &+\frac{2(\delta+\sqrt{N}\xi)}{\delta^2+\kappa^2+N\zeta^2+2\sqrt{N}\delta\zeta}\Bigg).
\end{align}
\end{subequations}
Using the initial state $|\Tilde{\Psi}_3(0)\rangle=|1_{k_1}1_{k_2};\textbf{0}_N;0\rangle$, the amplitudes in Eq.~(\ref{eq:photonwf3ls}) at time $t$ to lowest order in $J$ are given as
\begin{subequations}
\label{eq:3lsphsol}
\begin{align}
    \Tilde{n}_3(t)=&\sum^8_{j=1}\frac{\exp\left(-i\epsilon^{(j)}_3t\right)}{\prod_{k\neq j}\left(\epsilon^{(j)}_3-\epsilon^{(k)}_3\right)}\left(2i\kappa-2\omega_a+\epsilon^{(j)}_3\right)\nonumber\\
    &\times\left(\left(2\omega_a+\delta-i\kappa-\epsilon^{(j)}_3\right)^2-N\zeta^2\right)\nonumber\\
    &\times\left(\left(2\omega_a-\delta-i\kappa-\epsilon^{(j)}_3\right)^2-N\zeta^2\right)\nonumber\\
    &\times\left(\left(2\omega_a-2i\kappa-\epsilon^{(j)}_3\right)^2-4N\zeta^2\right),
\end{align}
\begin{align}
    \Tilde{p}_3(t)=&-\sqrt{N}J\sum^8_{j=1}\frac{\exp\left(-i\epsilon^{(j)}_3t\right)}{\prod_{k\neq j}\left(\epsilon^{(j)}_3-\epsilon^{(k)}_3\right)}\nonumber\\
    &\times\left(2i\kappa-2\omega_a+\epsilon^{(j)}_3\right)\left(2\omega_a+\delta-i\kappa-\epsilon^{(j)}_3\right)\nonumber\\
    &\times\left(\left(2\omega_a-\delta-i\kappa-\epsilon^{(j)}_3\right)^2-N\zeta^2\right)\nonumber\\
    &\times\left(\left(2\omega_a-2i\kappa-\epsilon^{(j)}_3\right)^2-4N\zeta^2\right),
\end{align}
\begin{align}
    \Tilde{q}_3(t)=&-\sqrt{N}J\sum^8_{j=1}\frac{\exp\left(-i\epsilon^{(j)}_3t\right)}{\prod_{k\neq j}\left(\epsilon^{(j)}_3-\epsilon^{(k)}_3\right)}\nonumber\\
    &\times\left(2i\kappa-2\omega_a+\epsilon^{(j)}_3\right)\left(2\omega_1-\delta-i\kappa-\epsilon^{(j)}_3\right)\nonumber\\
    &\times\left(\left(2\omega_a+\delta-i\kappa-\epsilon^{(j)}_3\right)^2-N\zeta^2\right)\nonumber\\
    &\times\left(\left(2\omega_a-2i\kappa-\epsilon^{(j)}_3\right)^2-4N\zeta^2\right),
\end{align}
\begin{align}
    &\Tilde{s}_3(t)=2\sqrt{2}NJ^2\sum^8_{j=1}\frac{\exp\left(-i\epsilon^{(j)}_3t\right)}{\prod_{k\neq j}\left(\epsilon^{(j)}_3-\epsilon^{(k)}_3\right)}\nonumber\\
    &\times\Bigg(2N\delta^2\zeta^2(2i\kappa-2\omega_a+\epsilon^{(j)}_3)+\nonumber\\
    &\left(\delta^2+N\zeta^2-\left(2\omega_a-i\kappa-\epsilon^{(j)}_3\right)^2\right)\Bigg(N\zeta^2\left(\epsilon^{(j)}_3-2\omega_a\right)\nonumber\\
    &+\left(2\omega_a-2i\kappa-\epsilon^{(j)}_3\right)\left(2\omega_a-i\kappa-\epsilon^{(j)}_3\right)\Bigg)\Bigg),
\end{align}
\begin{align}
\label{eq:3lsamp1}
    \Tilde{u}_3(t)=&NJ\zeta\sum^8_{j=1}\frac{\exp\left(-i\epsilon^{(j)}_3t\right)}{\prod_{k\neq j}\left(\epsilon^{(j)}_3-\epsilon^{(k)}_3\right)}\left(2i\kappa-2\omega_a+\epsilon^{(j)}_3\right)\nonumber\\
    &\times\left(\left(2\omega_a-\delta-i\kappa-\epsilon^{(j)}_3\right)^2-N\zeta^2\right)\nonumber\\
    &\times\left(\left(2\omega_a-2i\kappa-\epsilon^{(j)}_3\right)^2-4N\zeta^2\right),
\end{align}
\begin{align}
    \Tilde{v}_3(t)=&NJ\zeta\sum^8_{j=1}\frac{\exp\left(-i\epsilon^{(j)}_3t\right)}{\prod_{k\neq j}\left(\epsilon^{(j)}_3-\epsilon^{(k)}_3\right)}\left(2i\kappa-2\omega_a+\epsilon^{(j)}_3\right)\nonumber\\
    &\times\left(\left(2\omega_a+\delta-i\kappa-\epsilon^{(j)}_3\right)^2-N\zeta^2\right)\nonumber\\
    &\times\left(\left(2\omega_a-2i\kappa-\epsilon^{(j)}_3\right)^2-4N\zeta^2\right),
\end{align}
\begin{align}
    &\Tilde{w}_3(t)=2\sqrt{N^3}J^2\zeta\sum^8_{j=1}\frac{\exp\left(-i\epsilon^{(j)}_3t\right)}{\prod_{k\neq j}\left(\epsilon^{(j)}_3-\epsilon^{(k)}_3\right)}\nonumber\\
    &\Bigg(\delta^2\left(\epsilon^{(j)}_3-2\omega_a\right)-\left(6\omega_a-4i\kappa-3\epsilon^{(j)}_3\right)\times\nonumber\\
    &\left(N\zeta^2-\left(2\omega_a-i\kappa-\epsilon^{(j)}_3\right)^2\right)\Bigg)\left(2\omega_a-2i\kappa-\epsilon^{(j)}_3\right),
\end{align}
\begin{align}
\label{eq:3lsamp2}
    \Tilde{x}_3(t)=&-2\sqrt{2}N^2J^2\zeta^2\sum^8_{j=1}\frac{\exp\left(-i\epsilon^{(j)}_3t\right)}{\prod_{k\neq j}\left(\epsilon^{(j)}_3-\epsilon^{(k)}_3\right)}\nonumber\\
    &\times\Bigg(\delta^2\left(\epsilon^{(j)}_3-2\omega_a\right)-\left(6\omega_a-4i\kappa-3\epsilon^{(j)}_3\right)\nonumber\\
    &\times\left(N\zeta^2-\left(2\omega_a-i\kappa-\epsilon^{(j)}_3\right)^2\right)\Bigg).
\end{align}
\end{subequations}
The probabilities that the acceptor is singly- and doubly-excited, $\text{P}^{(1)}_3(t)$ and $\text{P}^{(2)}_3(t)$, respectively, are as follows
\begin{subequations}
\begin{eqnarray}
\label{eq:3lsprobss}
    \text{P}^{(1)}_3(t)=|\Tilde{u}_3(t)|^2+|\Tilde{v}_3(t)|^2+|\Tilde{w}_3(t)|^2,
\end{eqnarray}   
\begin{eqnarray}
\label{eq:3lsprobssb}
     \text{P}^{(2)}_3(t)=|\Tilde{x}_3(t)|^2.
\end{eqnarray}
\end{subequations}
Similar to the 2-LS acceptor, the long term dynamics are dominated by the terms with $\exp(-i\epsilon_3^{(1)}t)$ in Eqs.~(\ref{eq:3lsamp1})-(\ref{eq:3lsamp2}). The probabilities in Eqs.~(\ref{eq:3lsprobss}) and (\ref{eq:3lsprobssb}) can therefore be approximated as
\begin{subequations}
\begin{eqnarray}
    |\Tilde{u}_3(t)|^2=|\Tilde{v}_3(t)|^2\approx\frac{N^2J^2\zeta^2}{F_3}\exp\left(-\frac{E_3}{F_3}t\right),
\end{eqnarray}
\begin{eqnarray}
    |\Tilde{w}_3(t)|^2\approx\frac{4N^3J^4\zeta^2\kappa^2}{F_3^2}\exp\left(-\frac{E_3}{F_3}t\right),
\end{eqnarray}
\begin{eqnarray}
    |\Tilde{x}_3(t)|^2\approx\frac{2N^4J^4\zeta^4}{F_3^2}\exp\left(-\frac{E_3}{F_3}t\right),
\end{eqnarray}
\end{subequations}
where
\begin{eqnarray}
    &&E_3=4NJ^2 \kappa(\delta^2+\kappa^2+N\zeta^2),\nonumber\\
    &&F_3=(\delta^4+2\delta^2(\kappa^2-N\zeta^2)+(\kappa^2+N\zeta^2)^2).
\end{eqnarray}
To lowest order in $J$, $\text{P}^{(1)}_2(t)$ and $\text{P}^{(3)}_2(t)$ can be approximated as
\begin{subequations}
\begin{align}
\label{eq:ph3lsprob}
    \text{P}^{(1)}_3(t)=&\frac{2N^2J^2\zeta^2}{F_3}\exp\left(-\frac{E_3}{F_3}t\right),
\end{align}
\begin{align}
\label{eq:ph3lsprob2}
    \text{P}^{(2)}_3(t)=&\frac{2N^4J^4\zeta^4}{F_3^2}\exp\left(-\frac{E_3}{F_3}t\right).
\end{align}
\end{subequations}
The resulting one photon and two photon efficiencies, $\eta_3^{(1)}$ and $\eta_3^{(2)}$, are therefore given as
\begin{subequations}
\label{eq:eta3}
\begin{align}
\label{eq:eta31}
    \eta_3^{(1)}\approx\frac{N\zeta^2}{\delta^2+\kappa^2+N\zeta^2},
\end{align}
\begin{align}
\label{eq:eta32}
    \eta_3^{(2)}\approx\frac{2N^3J^2\zeta^4}{(\delta^2+\kappa^2+N\zeta^2)F_3}.
\end{align}
\end{subequations}

\section{Results}
Graphs of probabilities and efficiencies are plotted using the following parameter values: $\omega_a=12\xi$, $\kappa=\Gamma=0.3\xi$, $\zeta=\xi$, $J=0.01\xi$, and $\xi=10$ ps$^{-1}$. Probabilities are plotted against the dimensionless time parameters $\xi t$ for the 2-LS acceptor and $\zeta t$ for the 3-LS acceptor, and efficiencies are plotted against the dimensionless parameters $(2g+\Delta)/\xi$ and $J/\xi$ for the 2-LS acceptor, and $(2g+\Delta)/\zeta$, $(2g+\Delta)/\delta$,  and $J/\zeta$ for the 3-LS acceptor. 

\subsection{Probabilities}
\begin{figure}[h]
\includegraphics[width=0.48\textwidth]{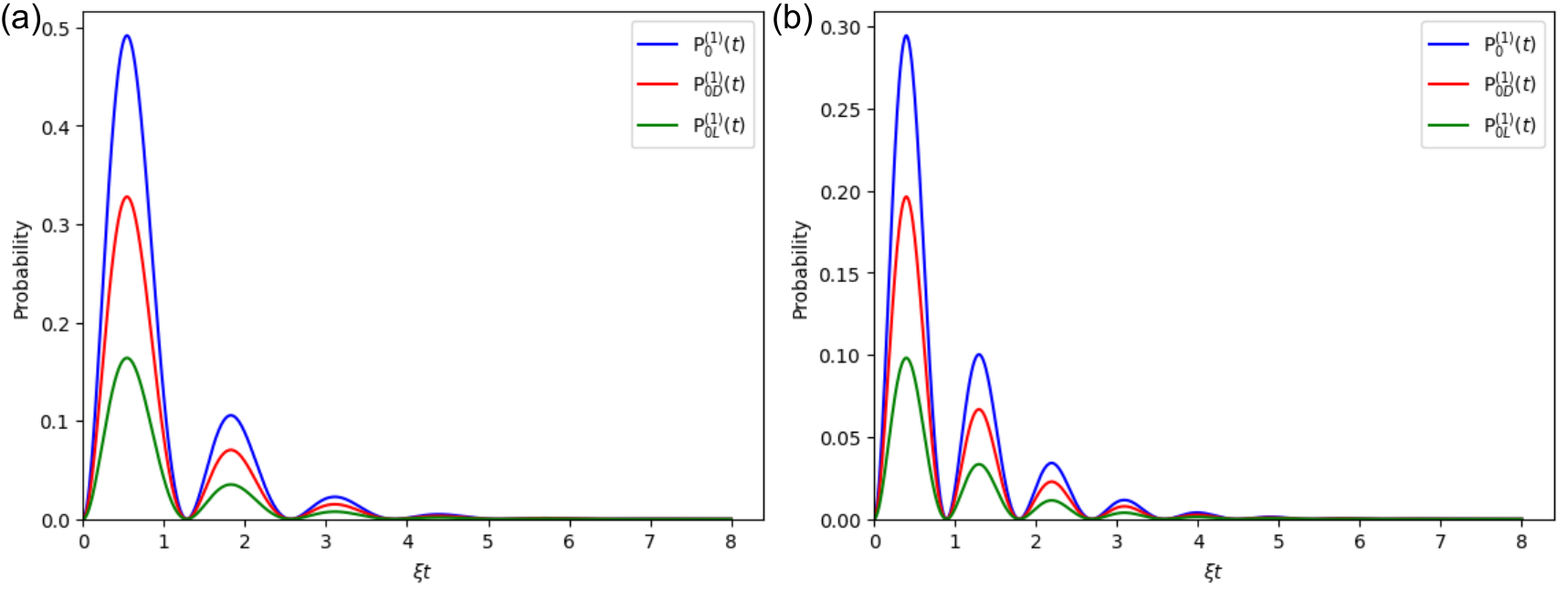}
\caption{\label{fig:prob0} Plots of the probabilities $\text{P}^{(1)}_0(t)$ (blue), $\text{P}^{(1)}_{0D}(t)$ (red) and $\text{P}^{(1)}_{0L}(t)$ (green) against $\xi t$ for a singly-excited 2-LS acceptor for states $|\Tilde{\Psi}_{0}(t)\rangle$ (with $N=3$), $|\Psi_{0D}(t)\rangle$, and $|\Psi_{0L}(t)\rangle$ in Eqs.~(\ref{eq:p10}), (\ref{eq:p10d}), and (\ref{eq:p10l}), respectively. In Fig.~2(a) $2g+\Delta=0$ and in Fig.~2(b) $2g+\Delta=5\xi$.}
\end{figure}

Probabilities $\text{P}^{(1)}_0(t)$, $\text{P}^{(1)}_{0D}(t)$ and $\text{P}^{(1)}_{0L}(t)$ in Eq.~(\ref{eq:p10}), (\ref{eq:p10d}), and (\ref{eq:p10l}), respectively, for a singly-excited 2-LS acceptor are plotted in Fig.~\ref{fig:prob0}. $\text{P}^{(1)}_0(t)$ is the acceptor single excitation probability corresponding to an initial double excitation on the ring with $N=3$ donor atoms. In Fig.~\ref{fig:prob0}(a) the energy levels of the collective donor ring and acceptor atom are on resonance, i.e. $\Delta+2g=0$, where $\Delta$ is the detuning between the acceptor and a single donor atom, and $g$ is the inter-donor coupling constant. In Fig.~\ref{fig:prob0}(b) the energy levels of the collective donor ring and acceptor are off-resonance, with $\Delta+2g=5\xi$. In Fig.~\ref{fig:prob0}(a) and Fig.~\ref{fig:prob0}(b), $\text{P}^{(1)}_0(t)$ exhibits a decaying oscillatory behavior, with periods $2\pi/(\sqrt{24}\xi)$ and $2\pi/(\sqrt{49}\xi)$, respectively and decay rate $0.6\xi$. It should be noted that the period in the on-resonance case is larger by a factor of approximately $\sqrt{2}$ compared to the off-resonance case. The initial probability maximum in the on-resonant case is $0.491$, which is larger by a factor of approximately $1.67$ compared to the off-resonance maximum of $0.294$. In both on- and off-resonance cases, the singly-excited acceptor probabilities, $\text{P}^{(1)}_{0D}(t)$ and $\text{P}^{(1)}_{0L}(t)$, for delocalized and localized initial states, respectively, have the same period and decay rate as $\text{P}^{(1)}_0(t)$. At all times $\text{P}^{(1)}_{0D}(t)=\tfrac{2}{3}\text{P}^{(1)}_0(t)$ and $\text{P}^{(1)}_{0L}(t)=\tfrac{1}{3}\text{P}^{(1)}_0(t)$. An initial state $|2_{N};0\rangle$ 
defined in Eq.~(\ref{eq:highnbasis}) simplifies when $N=3$ to
\begin{eqnarray}
    |\Psi_{0}(0)\rangle=\alpha|\Psi_{0D}(0)\rangle+\beta|\Psi_{0L}(0)\rangle,
\end{eqnarray}
where $\alpha=\sqrt{2/3}$ and $\beta=\sqrt{1/3}$. It should be noted that this initial state gives $\text{P}^{(1)}_0(t)$, which is the highest singly-excited acceptor probability at all times $t$. All other normalized values of $\alpha$ and $\beta$ give lower probabilities.

\begin{figure}[h]
\includegraphics[width=0.48\textwidth]{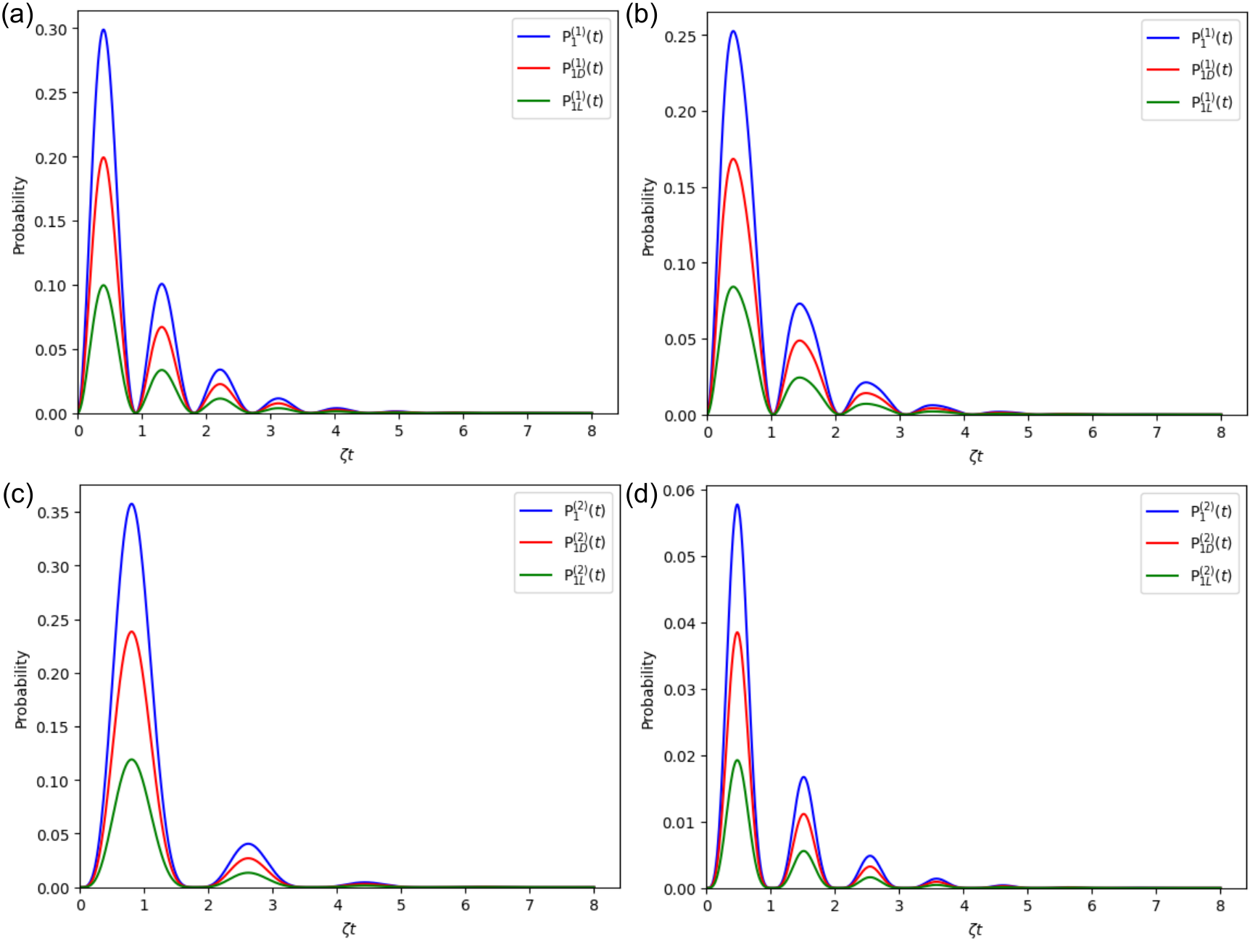}
\caption{\label{fig:prob1} (a) and (b) are plots of $\text{P}^{(1)}_1(t)$ (blue), $\text{P}^{(1)}_{1D}(t)$ (red) and $\text{P}^{(1)}_{1L}(t)$ (green) in Eq.~(\ref{eq:p1n}), (\ref{eq:p1d1}), and (\ref{eq:p1l1}), respectively, against $\zeta t$ for a singly-excited 3-LS acceptor. (c) and (d) are plots of $\text{P}^{(2)}_1(t)$ (blue), $\text{P}^{(2)}_{1D}(t)$ (red) and $\text{P}^{(2)}_{1L}(t)$ (green) from Eq.~(\ref{eq:ps1n}), (\ref{eq:p1d2}), and (\ref{eq:p1l2}), respectively, against $\zeta t$ for a doubly-excited 3-LS acceptor. In Figs.~3(a) and 3(c) $2g+\Delta=0$ and in Figs.~3(b) and 3(d) $2g+\Delta=5\zeta$.}
\end{figure}
The probabilities $\text{P}^{(1)}_1(t)$, $\text{P}^{(1)}_{1D}(t)$ and $\text{P}^{(1)}_{1L}(t)$ in Eqs.~(\ref{eq:p1n}), (\ref{eq:p1d1}), and (\ref{eq:p1l1}), for a singly-excited 3-LS acceptor are plotted in Fig.~\ref{fig:prob1}(a) and Fig.~\ref{fig:prob1}(b) for the on-resonance ($2g+\Delta=0$) and off-resonance ($2g+\Delta=5\zeta$) cases, respectively. Similar to the 2-LS acceptor, all probabilities exhibit a decaying oscillatory behavior. The on-resonance single excitation probability, $\text{P}^{(1)}_1(t)$, has a period of $\pi/(\sqrt{12}\zeta)$ while the off-resonance period is $2\pi/(\sqrt{37}\zeta)$. Compared to the 2-LS acceptor single excitation probability, the on-resonance period for the 3-LS acceptor is smaller by a factor of $0.707$, while the off-resonance period is larger by a factor of $1.15$. The initial probability maximum for the on-resonance case is $0.298$, and off-resonance it is $0.252$. Both these probabilities are smaller compared to the corresponding probabilities for the 2-LS acceptor. The single excitation probabilities, $\text{P}^{(1)}_{1D}(t)$, and $\text{P}^{(1)}_{1L}(t)$, for the delocalized and localized initial states, respectively, have the same period, $\pi/(\sqrt{12}\zeta)$ for the on-resonance case and $\pi/(\sqrt{37}\zeta)$ for the off-resonance case. $\text{P}^{(1)}_{1D}(t)$ and $\text{P}^{(1)}_{1L}(t)$ have initial maxima of $0.198$ and $0.0993$ in the on-resonance case, and $0.168$ and $0.0840$ in the off-resonance case, respectively. It should be noted that the initial maximum of $\text{P}^{(1)}_{1D}(t)$ and $\text{P}^{(1)}_{1L}(t)$ is 2/3 and 1/3 of the initial maximum of $\text{P}^{(1)}_{1}(t)$, respectively. The periods of $\text{P}^{(1)}_{1D}(t)$ and $\text{P}^{(1)}_{1L}(t)$ are the equal to the period of $\text{P}^{(1)}_{1}(t)$.

We now consider $\text{P}^{(2)}_1(t)$, $\text{P}^{(2)}_{1D}(t)$ and $\text{P}^{(2)}_{1L}(t)$ in Eqs.~(\ref{eq:ps1n}), (\ref{eq:p1d2}), and (\ref{eq:p1l2}), respectively, for the doubly-excited 3-LS acceptor in Fig.~\ref{fig:prob1}(c) (on-resonance) and Fig.~\ref{fig:prob1}(d) (off-resonance). The period of $\text{P}^{(2)}_1(t)$ in the on-resonance case is $2\pi/(\sqrt{12}\zeta)$, which is twice the period of $\text{P}^{(1)}_1(t)$. In general, this means that on-resonance there is faster transfer of single excitations compared to double excitations. However, in the off-resonance case the period $2\pi/(\sqrt{37}\zeta)$, which is the same as that for $\text{P}^{(1)}_1(t)$. This indicates that the transfer time for both single and double excitations is the same in the off-resonance case. The fastest transfer occurs for single excitations in the on-resonance case, while the slowest transfer is for on-resonance double excitations. The initial maximum for $\text{P}^{(1)}_2(t)$ on-resonance is $0.358$, while the off-resonance value is $0.058$. The double excitation probabilities for the delocalized initial state, $\text{P}^{(2)}_{1D}(t)$, and localized initial state, $\text{P}^{(2)}_{1L}(t)$, have the same period of $2\pi/(\sqrt{12}\zeta)$ and $2\pi/(\sqrt{37}\zeta)$ for the on- and off-resonance cases, respectively. The initial maxima of $\text{P}^{(2)}_{1D}(t)$ and $\text{P}^{(2)}_{1L}(t)$ are $0.238$ and $0.039$ for on-resonance, and $0.119$ and $0.019$ for off-resonance, respectively. Similar to the single excitation case, it should be noted the initial maximum of $\text{P}^{(2)}_{1D}(t)$ and $\text{P}^{(2)}_{1L}(t)$ is 2/3 and 1/3 of the initial maximum of $\text{P}^{(2)}_{1}(t)$, respectively. Also, the periods of $\text{P}^{(2)}_{1D}(t)$ and $\text{P}^{(2)}_{1L}(t)$ are the equal to the period of $\text{P}^{(2)}_{1}(t)$.

\begin{figure}[h]
\includegraphics[width=0.48\textwidth]{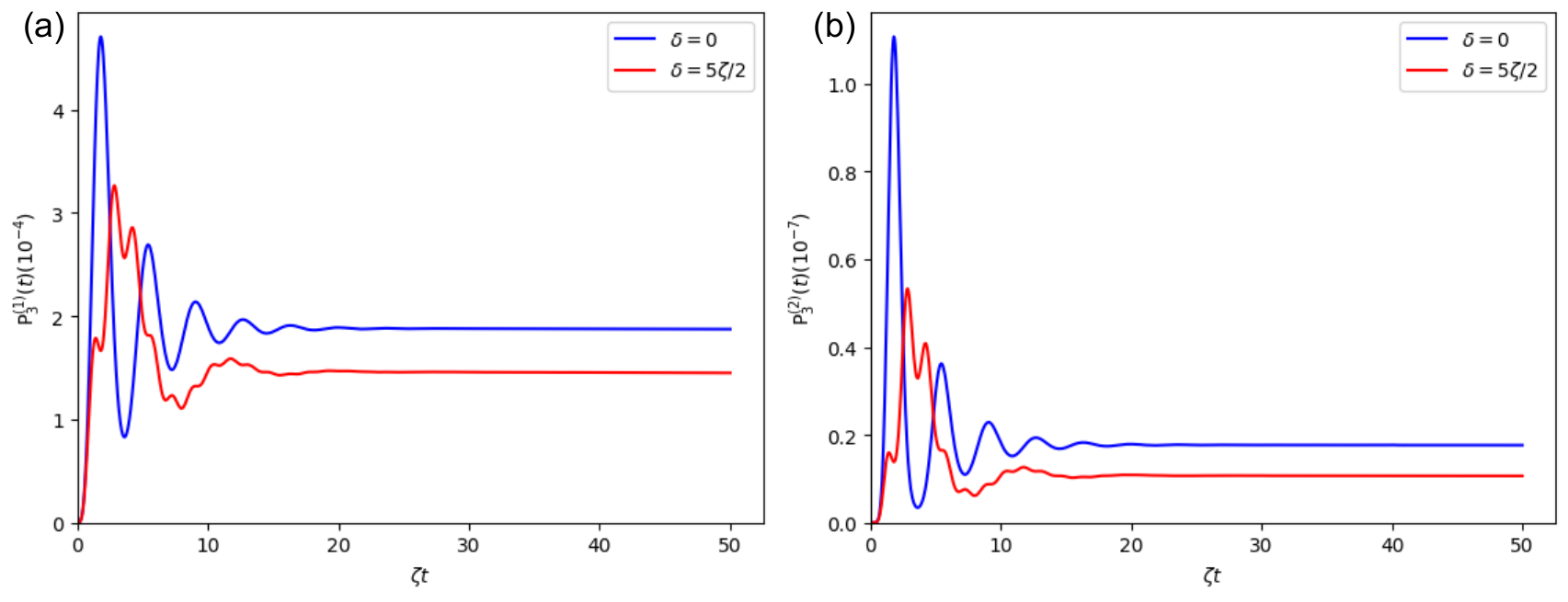}
\caption{\label{fig:prob2} Plots of probabilities (a) $\text{P}^{(1)}_{3}(t)$ and (b) $\text{P}^{(2)}_{3}(t)$ against the $\zeta t$ in Eq.~(\ref{eq:ph3lsprob}) for a 3-LS acceptor in the singly- and doubly-excited state, respectively, for $N=3$ and an incident photon pair. Both plots correspond to $\delta=0$ (blue) and $\delta=5\zeta/2$ (red) for the on-resonance and off-resonance cases, respectively.}
\end{figure}

Fig.~\ref{fig:prob2} gives the probabilities $\text{P}^{(1)}_{3}(t)$ and $\text{P}^{(2)}_{3}(t)$ against $\zeta t$ in Eqs.~(\ref{eq:ph3lsprob}) and~(\ref{eq:ph3lsprob2}) with $N=3$ and an incident photon pair, for the 3-LS acceptor in a singly- and doubly-excited state, respectively. It should be noted that $\text{P}^{(1)}_{2}(t)=\text{P}^{(1)}_{3}(t)$, where $\text{P}^{(1)}_{2}(t)$ is the singly-excited probability for the 2-LS acceptor with an incident photon pair and therefore $\text{P}^{(1)}_{3}(t)$ also describes $\text{P}^{(1)}_{2}(t)$. Fig.~\ref{fig:prob2}(a) shows the plots $\text{P}^{(1)}_{3}(t)$, for the singly-excited 3-LS acceptor, for the $\delta=0$ (blue) and for $\delta=5\zeta/2$ (red) cases. In both cases, the probabilities exhibit damped oscillations, which asymptotically decay to $1.89\times10^{-4}\exp(-1.17\times10^{-4}\zeta t)$ in the on-resonance case and  $1.47\times10^{-4}\exp(-2.75\times10^{-4}\zeta t)$ in the off-resonance case, for $\zeta t\gtrapprox25$. Fig.~\ref{fig:prob2}(b) shows the plots of the probabilities, $\text{P}^{(2)}_{3}(t)$, for the doubly-excited 3-LS acceptor in the on-resonance and off-resonance cases. Similar to the singly-excited 3-LS acceptor, both probabilities exhibit initial damped oscillations, which asymptotically decay to $1.78\times10^{-8}\exp(-1.17\times10^{-4}\zeta t)$ in on-resonance case, and to $1.08\times10^{-8}\exp(-2.75\times10^{-4}\zeta t)$ in the off-resonance case. $\text{P}^{(1)}_{3}(t)$ and $\text{P}^{(2)}_{3}(t)$ have initial maxima of the order of $10^{-4}$ and $10^{-7}$, respectively, while the probabilities without light, $\text{P}^{(1)}_{1}(t)$ and $\text{P}^{(2)}_{1}(t)$, both have initial maxima of the order of $10^{-1}$.

\subsection{Transfer Efficiencies}
\begin{figure}[h]
\includegraphics[width=0.48\textwidth]{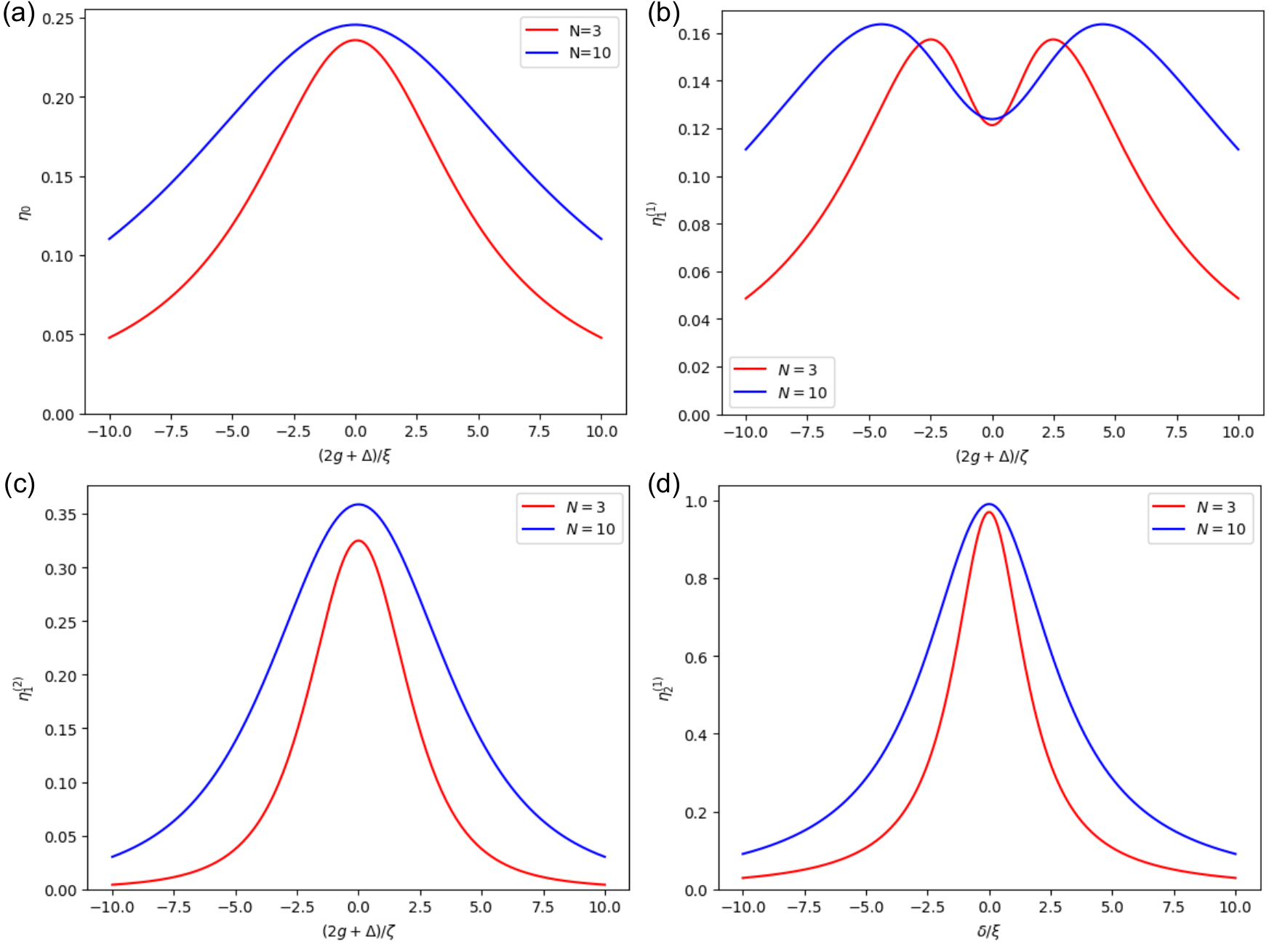}
\caption{\label{fig:eta}(a) Plots of the single excitation efficiency $\eta^{(1)}_{0}$ in Eq.~(\ref{eq:eta0n}) for 2-LS acceptor against the dimensionless detuning $(2g+\Delta)/\xi$. (b) Plots of the single excitation efficiency $\eta^{(1)}_1$ in Eq.~(\ref{eq:etat}) for a 3-LS acceptor against the dimensionless detuning $(2g+\Delta)/\zeta$. (c) Plots of the double excitation efficiency $\eta^{(2)}_1$ in Eq.~(\ref{eq:eta1}) for a 3-LS acceptor against the dimensionless detuning $(2g+\Delta)/\zeta$. (d) Plots of the single excitation efficiency $\eta^{(1)}_2$ in Eq.~(\ref{eq:eta12}) for 2-LS acceptor with an incident photon pair against the dimensionless detuning $\delta/\xi$. The graphs plotted are for $N=3$ (red) and $N=10$ (blue).}
\end{figure}
\begin{figure*}
\includegraphics[width=0.96\textwidth]{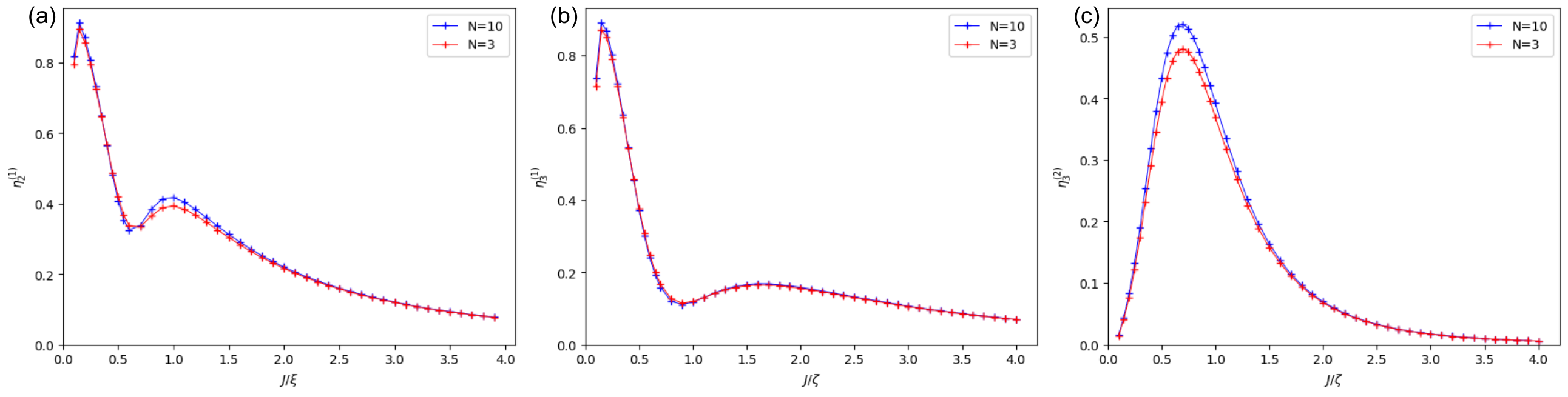}
\caption{\label{fig:eta2} Numerical plots in the non-perturbative regime for (a) single excitation efficiency $\eta^{(1)}_2$ against $J/\xi$ for a 2-LS acceptor, (b) single excitation efficiency $\eta^{(1)}_3$ against $J/\zeta$ for a 3-LS acceptor, (c) double excitation efficiency $\eta^{(2)}_3$ against $J/\zeta$ for a 3-LS acceptor with an incident photon pair. Plots correspond to $N=3$ (red) and $N=10$ (blue), for $\delta=0$.}
\end{figure*}
Fig.~\ref{fig:eta} shows the plots of the efficiencies  $\eta^{(1)}_0$ (Eq.~(\ref{eq:eta0n})), $\eta^{(1)}_1$ (Eq.~(\ref{eq:etat})), $\eta^{(2)}_1$ (Eq.~(\ref{eq:eta1})), and $\eta^{(1)}_2$ (Eq.~(\ref{eq:eta12})) against the detuning parameters $(2g+\Delta)/\xi$, $(2g+\Delta)/\zeta$, and $\delta/\xi$. Fig.~\ref{fig:eta}$(a)$ gives the plots of the single excitation efficiency $\eta^{(1)}_0$ in Eq.~(\ref{eq:eta0n}) for the 2-LS acceptor against $(2g+\Delta)/\xi$, for $N=3$ and $N=10$. The maximum efficiencies for $\eta^{(1)}_0$ occur at zero detuning and are $0.236$ for $N=3$ and $0.246$ for $N=10$. The maximum efficiency for arbitrary $N$ is given as $\eta^{(1)}_0=\frac{N}{0.72+4N}$. The dimensionless bandwidth of $\eta^{(1)}_0$ for $N=3$ it is $6.95$ while for $N=10$ is $12.66$. It should be noted that the efficiency for $N=10$ is always greater than the efficiency for $N=3$, for all values of the dimensionless detunings.

Fig.~\ref{fig:eta}$(b)$ gives the plots of the single excitation efficiency $\eta^{(1)}_1$ in Eq.~(\ref{eq:etat}) for the 3-LS acceptor against $(2g+\Delta)/\zeta$. In this case the maximum efficiency for $N=3$ is $0.157$ occurring at dimensionless detuning values $\pm4.49$ and for $N=10$, the maximum is $0.164$ occurring at $\pm2.48$. The dimensionless bandwidth of $\eta^{(1)}_1$ for $N=3$ is $14.45$ while for $N=10$ it is $25.89$. It should be noted that $\eta^{(1)}_1$ has a minimum of $12.1$ for $N=3$ and $12.4$ for $N=10$ on-resonance.

Fig.~\ref{fig:eta}$(c)$ gives the plots of the double excitation efficiency $\eta^{(2)}_1$ in Eq.~(\ref{eq:eta1}) for the 3-LS acceptor against $(2g+\Delta)/\zeta$. The maximum efficiencies of $\eta^{(1)}_1$ occur at zero detuning and are $0.325$ for $N=3$ and $0.359$ for $N=10$. The maximum efficiency for arbitrary $N$ is given as $\eta^{(2)}_1=\frac{6N^2}{(4N+\frac{9}{25})(4N+\frac{36}{25})}$. The bandwidth of $\eta^{(2)}_1$ for $N=3$ is $4.62$ while for $N=10$ it is $8.23$. Similar to $\eta^{(1)}_0$ the efficiency for $N=10$ is always greater than the efficiency for $N=3$.

Fig.~\ref{fig:eta}$(d)$ gives the plots of the single excitation efficiency $\eta^{(1)}_2$ in Eq.~(\ref{eq:eta12}) for two incident photons for the 2-LS acceptor against $\delta/\xi$. The maximum efficiencies of $\eta^{(1)}_2$ occur at zero detuning and are $0.971$ for $N=3$ and $0.991$ for $N=10$. The maximum efficiency for arbitrary $N$ is given as $\eta^{(1)}_2=\frac{N}{\frac{9}{100}+N}$. The bandwidth of $\eta^{(1)}_2$ for $N=3$ is $3.51$ while for $N=10$ it is $6.32$. It should be noted that the efficiency for $N=10$ is always greater than the efficiency for $N=3$. In addition, the single excitation efficiency $\eta^{(1)}_2$ in Eq.~(\ref{eq:eta12}) for the 2-LS acceptor is the same as the single excitation efficiency for the 3-LS acceptor, $\eta^{(1)}_3$, in Eq.~(\ref{eq:eta31}), for $\zeta=\xi$. It is found the double excitation transfer efficiency $\eta^{(2)}_3$ is negligible in the perturbative regime $J<<\zeta$.

Fig.~\ref{fig:eta2}$(a)$ gives numerical plots in the non-perturbative regime for the single excitation efficiency $\eta^{(1)}_2$ against $J/\xi$ for a 2-LS acceptor with an incident on-resonant photon pair for $N=3$ and $N=10$. Maxima of $0.895$ at $J=0.15\xi$ for $N=3$ and $0.913$ at $J=0.15\xi$ for $N=10$ are obtained. In contrast to the perturbative regime in Fig.~\ref{fig:eta}$(d)$, the simulated efficiencies are almost identical for $N=3$ and $N=10$. However, in perturbative regime, with $J=0.01\xi$, near-unity maxima are obtained for $N=3$ and $N=10$. Fig.~\ref{fig:eta2}$(b)$ gives plots for the single excitation efficiency $\eta^{(1)}_3$ against $J/\zeta$ for a 3-LS acceptor with an incident on-resonant photon pair for $N=3$ and $N=10$. Maxima of $0.870$ at $J=0.15\zeta$ for $N=3$ and $0.890$ at $J=0.15\zeta$ for $N=10$ are obtained. The single excitation transfer efficiency is higher for the 2-LS acceptor for all values of the dimensionless coupling $J/\zeta$ compared to that of the 3-LS acceptor. It should be noted that there is a sharp decrease in the efficiency $\eta^{(1)}_3$ for couplings $0.15\geq J/\zeta<0.9$, followed by a slight increase between $0.9\geq J/\zeta<1.6$ and a general decrease for $J/\zeta>1.6$. Fig.~\ref{fig:eta2}$(c)$ gives plots for the double excitation efficiency $\eta^{(2)}_3$ against $J/\zeta$ for a 3-LS acceptor with an incident on-resonant photon pair for $N=3$ and $N=10$. Maxima of $0.481$ at $J=0.70\zeta$ for $N=3$ and $0.521$ at $J=0.70\zeta$ for $N=10$ are obtained. It is observed that larger photon couplings are required for maximum double excitation transfer efficiency compared to the single excitation transfer efficiency. However, the maximum double excitation transfer efficiencies are smaller and approximately $60\%$ of the single excitation transfer efficiencies. Additional numerical simulations with an incident off-resonant photon pair resulted in reduced transfer efficiencies in all cases.

\section{Conclusion}
In this paper we derived analytically and simulated numerically the acceptor probabilities and transfer efficiencies for a ring antenna light harvesting system coupled to either a 2-LS or a 3-LS acceptor atom with no light or with an incident photon pair. All acceptor probabilities in the no light case with an initial double excitation on the ring exhibit a decaying oscillatory behavior over a time scale of approximately $40$ps. Generally, a less than $50\%$ efficiency was obtained in both the single excitation transfer and double excitation transfer cases. 

In the perturbative limit ($J<<\xi$) with an incident photon pair, single excitation probabilities for a 2-LS or a 3-LS acceptor are 3 orders of magnitude smaller than in the no-light case and exhibit short term oscillatory behavior as well as long term decay over a time scale of hundreds of nanoseconds. Near-perfect efficiency was obtained for single excitation transfer with on-resonant photon pairs. Double excitation probabilities were a factor of 1000 smaller than single excitation probabilities but with the same decay time scale leading to negligible double excitation transfer efficiency.

In the non-perturbative case with an incident on-resonance photon pair, numerical simulations demonstrated an over $90\%$ single excitation transfer efficiency and approximately $50\%$ double excitation transfer efficiency. Off-resonant photon pairs tend to be transferred less efficiently compared to on-resonant photon pairs.

In conclusion, the three factors that determine high excitation transfer efficiency in LH1-RC type photosynthetic units are relatively high acceptor probabilities, long decay times, and for double excitation transfer, strong photon-ring coupling. The theoretical framework presented in this paper can be implemented in the practical design of light-harvesting technologies which will be useful for the realization of highly efficient bio-inspired solar energy devices. \cb 
However, translating these findings into a practical device requires addressing several key considerations. Below we outline a possible experimental strategy and discuss the materials and engineering challenges, drawing upon recent advances in the field.

\subsection{Device Architecture and Materials}
A feasible practical implementation would adopt a bio-inspired architecture featuring a donor ring coupled to a central acceptor, mimicking the LH1-RC arrangement in natural photosynthetic bacteria. In practice, the donor ring may be realized using one of the following approaches:
\begin{itemize}
    \item \textbf{Self-assembled Nanostructures:} Organic dye molecules or semiconductor quantum dots can be functionalized with ligands to promote self-assembly into ring structures. The close-packing and precise spatial arrangement are critical for strong inter-donor coupling, which is essential to achieve the high transfer efficiencies predicted by our model.
    \item \textbf{Nanofabrication Techniques:} Advanced lithographic methods or templated deposition (e.g., on pre-patterned substrates) may be used to engineer the precise geometry of the ring. Such techniques allow for fine tuning of donor–donor spacing and coupling constants, thereby optimizing the photon–ring interaction.
\end{itemize}

For the central acceptor:
\begin{itemize}
    \item \textbf{Molecular Acceptors:} A robust candidate would be a specially designed molecular complex with a two-level or three-level electronic structure, engineered to have a long-lived excited state. This could be achieved through modifications on conventional porphyrin or phthalocyanine systems.
    \item \textbf{Hybrid Materials:} Integration of organic and inorganic materials—such as combining conjugated polymers with metal oxides or perovskite nanostructures—can lead to acceptors with tailored electronic properties and enhanced stability.
\end{itemize}

\subsection{Integration with Electrodes and Photoelectrochemical Cells}
For energy conversion applications, the self-assembled LH complex can be integrated into a photoelectrochemical cell:
\begin{itemize}
    \item \textbf{Electrode Immobilization:} Following the approach of Suemori \emph{et al.} \cite{exp1}, the donor-acceptor assemblies can be immobilized onto transparent conductive substrates such as indium tin oxide (ITO) electrodes. Such arrangements have been shown to facilitate efficient photocurrent generation by providing a direct path for charge collection.
    \item \textbf{Charge Extraction Layers:} In addition to the conducting electrode, proper electron and hole extraction layers should be designed. These could be realized with metal oxides (e.g., TiO$_2$ for electrons and NiO for holes) that are widely used in dye-sensitized and perovskite solar cells.
    \item \textbf{Encapsulation and Stability:} Stability is a critical issue in practical devices. Encapsulation strategies, for instance using polymer matrices or robust inorganic coatings, can protect the delicate donor and acceptor components from environmental degradation (e.g., moisture and oxygen).
\end{itemize}

\subsection{Parameter Optimization Based on Analytical Insights}
Our analytical results suggest that the overall device performance will be maximized by:
\begin{enumerate}
    \item \textbf{Maximizing the Acceptor’s Excitation Probability:} This can be achieved by selecting acceptor materials with inherently high absorption cross-sections and engineering strong donor–acceptor coupling through chemical modification (e.g., tethering or covalent linking).
    \item \textbf{Prolonging the Excited State Lifetimes:} Materials with low non-radiative decay channels are preferred. This might involve the incorporation of heavy atoms or the design of rigid molecular frameworks that reduce vibrational losses.
    \item \textbf{Enhancing Photon–Ring Coupling:} Nanostructuring approaches such as plasmonic enhancement (using noble metal nanoparticles) or dielectric resonators can be explored to further amplify the local electromagnetic field, thereby increasing the effective coupling.
\end{enumerate}
By systematically tuning these parameters, one can tailor the quantum efficiency of the transfer process to meet the requirements of a scalable solar energy device.

\subsection{Scalability and Future Directions}
In terms of scalability, the self-assembly and nanofabrication techniques discussed are already being explored in the context of organic photovoltaics and artificial photosynthesis \citep{nocera2012,Gust2009,zhao2023}. Future research could focus on:
\begin{itemize}
    \item \textbf{Hybrid Integration:} Combining the light-harvesting unit with semiconductor devices to create integrated photoelectrochemical systems.
    \item \textbf{In-situ Characterization:} Advanced spectroscopic and microscopic techniques will be critical for monitoring the assembly and charge transfer dynamics in real time.
    \item \textbf{Optimization through Simulation:} Computational modeling can help to predict optimal configurations and guide the synthesis of new materials with tailored electronic properties.
\end{itemize}
Ultimately, the experimental implementation of these artificial light-harvesting devices, guided by our detailed analytical insights, holds great promise for the development of next-generation renewable energy technologies. \cbl

\begin{acknowledgments}
We would like to thanks C.A. Downing for helpful discussions. OIRF is funded by the EPSRC via the Maths DTP 2021-22 University of Exeter (EP/W523859/1).
\end{acknowledgments}

\appendix
\section{\label{appen1}Ring Operators}
The $j$-th 3-LS on the ring has states $|0_j\rangle$, $|1_j\rangle$, and $|2_j\rangle$. We have allowed transitions $|0_j\rangle\rightarrow|1_j\rangle$ ($|1_j\rangle\rightarrow|0_j\rangle$), defined by the operator $S^{j}_{10}$ ($S^{j}_{01}$), and $|1_j\rangle\rightarrow|2_j\rangle$ ($|2_j\rangle\rightarrow|1_j\rangle$), with the operator $S^{j}_{21}$ ($S^{j}_{12}$). In bra-ket notation, these operators are as follows
\begin{align}
    S^{j}_{10}=|1_j\rangle\langle0_j|,\quad\quad S^{j}_{01}=|0_j\rangle\langle1_j|,\\
    S^{j}_{21}=|2_j\rangle\langle1_j|,\quad\quad S^{j}_{12}=|1_j\rangle\langle2_j|.
\end{align}
Additionally, we introduce the operators $S^j_{ii}=|i\rangle\langle i|$. The Hamiltonian of the ring system with these operators is
\begin{align}
    H_d=&\sum_{j=1}^N\left[(\omega_d-i\kappa)S^j_{11}+2(\omega_d-i\kappa)S^j_{22}\right.\nonumber\\
    &\left.+g\left((S^{j}_{10}+\sqrt{2}S^{j}_{21})(S^{j+1}_{01}+\sqrt{2}S^{j+1}_{12})+h.c.\right)\right].
\end{align}
We have chosen couplings for transitions between the $|0\rangle$ and $|1\rangle$ states on both 3-LSs to be $g$, coupling between the $|0\rangle$ and $|1\rangle$ on one 3-LS and the $|1\rangle$ and $|2\rangle$ on the other to be $\sqrt{2}g$, and coupling between $|1\rangle$ and $|2\rangle$ on both 3-LSs to be $2g$~\cite{aiyejina}.

Introducing the creation (annihilation operators) $e^\dagger_j$ ($e_j$), defined in terms of transition operators
\begin{align}
    e^\dagger_j=S^j_{10}+\sqrt{2}S^j_{21}\quad\quad e_j=S^j_{01}+\sqrt{2}S^j_{12}.
\end{align}
These satisfy the site-dependent commutation relations
\begin{align}
    [e_i,e_j^\dagger]=\delta_{ij}(|0_i\rangle\langle0_j|+|1_i\rangle\langle1_j|-2|2_i\rangle\langle2_j|).
\end{align}
The Hamiltonian $H_d$ with these operators becomes
\begin{equation}
    H_d=\sum_{j=1}^N\left[(\omega_d-i\kappa)e_j^\dagger e_j+g(e^\dagger_je_{j+1}+h.c.)\right].
\end{equation}

\section{\label{appen}Collective State Operators}
The Fourier transformation of the ring operators in Eq.~(\ref{eq:ejdef}) is given as
\begin{subequations}
\begin{equation}
   e_j=\frac{1}{\sqrt{N}}\sum_{k}e^{ijk}\Tilde{\textbf{e}}_{k},
\end{equation}
\begin{equation}
    e^\dagger_j=\frac{1}{\sqrt{N}}\sum_{k}e^{-ijk}\Tilde{\textbf{e}}^\dagger_{k},
\end{equation}
\end{subequations}
where $\Tilde{\textbf{e}}^\dagger_{k}$ ($\Tilde{\textbf{e}}_{k}$) are the creation (annihilation) collective momentum-space operators defined by momentum $k$, with $k=\frac{2\pi(i-1)}{N}$ and $i=1,..,N$. The momentum-space operators have commutation relations
\begin{eqnarray}
    [\Tilde{\textbf{e}}_k,\Tilde{\textbf{e}}^\dagger_{k'}]=\delta_{kk'}\left(1-\frac{3}{N}\sum_{j=1}^N|2_j\rangle\langle2_j|\right).
\end{eqnarray}

The ring Hamiltonian, $H_d$ in Eq.~(\ref{eq:ringhamil}), with these new operators becomes
\begin{eqnarray}
\label{eq:hdkspace}
    \Tilde{H}_d=\sum_k(\omega_d-i\kappa+2g\cos(k))\Tilde{\textbf{e}}^\dagger_{k}\Tilde{\textbf{e}}_{k}.
\end{eqnarray}
Additionally, the donor-acceptor coupling, $H_{da}$ in Eq.~(\ref{eq:hda}), for the 2-LS acceptor becomes
\begin{align}
    \Tilde{H}_{da}=\xi(\Tilde{\textbf{e}}^\dagger_{0}a+\Tilde{\textbf{e}}_{0}a^\dagger),
\end{align}
since the only mode that couples to the acceptor is the $k=0$ mode.

\nocite{*}
\bibliography{apssamp}

\end{document}